\documentclass[aps,prb,groupedaddress,  superscriptaddress, twocolumn, notitlepage]{revtex4-2}
\usepackage[utf8]{inputenc}
\usepackage{amsmath}
\usepackage{amsfonts}
\usepackage{amssymb}
\usepackage{braket}
\usepackage{graphicx}
\usepackage{booktabs}
\usepackage[colorlinks=true, allcolors=blue]{hyperref}
\newcommand{\normord}[1]{{:}\!\mathrel{#1}\!{:}}

\begin{document}
\title{Model wavefunctions for an interface between lattice Laughlin and Moore-Read states}
\author{B\l a\.{z}ej Jaworowski}
%\altaffiliation{On leave from Department of Theoretical Physics, Wroc\l aw University of Science and Technology, PL-50370 Wroc\l aw, Poland}
\email{blazej@phys.au.dk}
\affiliation{Max Planck Institute for the Physics of Complex Systems, D-01187 Dresden, Germany}
\affiliation{Department of Physics and Astronomy, Aarhus University, DK-8000 Aarhus C, Denmark}
\author{Anne E. B. Nielsen}
\affiliation{Max Planck Institute for the Physics of Complex Systems, D-01187 Dresden, Germany}
\affiliation{Department of Physics and Astronomy, Aarhus University, DK-8000 Aarhus C, Denmark}
\begin{abstract}
We use conformal field theory to construct model wavefunctions for a gapless interface between lattice versions of a bosonic Laughlin state and a fermionic Moore-Read state, both at $\nu=1/2$. The properties of the resulting model state, such as particle density, correlation function and R\'enyi entanglement entropy are then studied using the Monte Carlo approach. Moreover, we construct the wavefunctions also for anyonic excitations (quasiparticles and quasiholes). We study their density profile, charge and statistics. We show that, similarly to the Laughlin-Laughlin case studied earlier, some anyons (the Laughlin Abelian ones) can cross the interface, while others (the non-Abelian ones) lose their anyonic character in such a process. Also, we argue that, under an assumption of local particle exchange, multiple interfaces give rise to a topological degeneracy, which can be interpreted as originating from Majorana zero modes.
\end{abstract}
\maketitle
\section{Introduction}
One of the characteristic features of the topological orders is the existence of nontrivial physical phenomena at the edges or interfaces with another topological phase. While the former can be used to characterize a single topological phase \cite{wen1992theory,heiblum2020edge}, the latter can tell us how two different topological phases are related to each other (e.g. if one of them can be transformed into the other by anyon condensation  \cite{bombin2008family,bais2009theory,bais2009condensate, bais2010topological, bombin2011nested, Beigi2011,Kitaev2012,bais2012modular,burnell2012phase, burnell2018anyon}). In experiments, the interfaces can be potentially useful for example for isolating a certain edge mode to prove its existence \cite{yang2017interface,crepel2019variational}. From the perspective of applications, the interfaces can exhibit additional topological degeneracy connected to the existence of non-Abelian zero-energy modes similar to the Majorana zero modes  \cite{bombin2011nested, lindner2012fractionalizing,clarke2013exotic, santos2017parafermionic,lian2018theory}, which can encode quantum logic gates \cite{hutter2016quantum,clarke2013exotic,mong2014universal,lindner2012fractionalizing}. Curiously, this can happen even if both sides are Abelian \cite{lindner2012fractionalizing,clarke2013exotic, santos2017parafermionic}.

As the fractional quantum Hall (FQH) \cite{laughlin,tsui} states are the paradigmatic examples of topological orders, the study of interfaces between them is particularly important. Some such interfaces can be created experimentally, e.g. the Camino et al.'s anyonic interferometry experiments used a system with two different filling factors \cite{camino2005realization}. Proposals for realizing other interfaces also do exist, involving either two Abelian states \cite{crepel2019microscopic,crepel2019model} or one Abelian and one non-Abelian state \cite{yang2017interface}. Moreover, one should bear in mind that FQH states can exist in lattice settings, as shown in a number of numerical studies \cite{Neupert,PRX,SunNature,Zoology} and experiments with graphene moir\'e superlattices subjected to magnetic fields \cite{spanton2018observation}. There are ongoing attempts to create lattice FQH states, including their bosonic versions, in optical lattices \cite{sorensen2005fractional, palmer2006high,palmer2008optical,kapit2010exact,moller2009composite,hafezi2007fractional,yao2013realizing,cooper2013reaching, nielsen2013local}. The large degree of control that we can exert on these systems make them potentially useful also for the study of the interfaces. 

%In contrast to interfaces between the nochiral topological orders \cite{bombin2011nested,Beigi2011,Kitaev2012}, we are not aware of any exactly solvable models for the FQH interfaces.
A majority of studies on the FQH interfaces focus on their most general features, using ``top-down'' techniques based on field theory: the topological symmetry breaking formalism (i.e. anyon condensation) \cite{bais2009theory,bais2009condensate, bais2010topological, burnell2012phase,bais2012modular, burnell2018anyon}, coupling of effective edge theories \cite{cano2015interactions, barkeshli2015particle,wan2016striped,yang2017interface,santos2018symmetry,mross2018theory,lian2018theory,mross2018theory, wang2018topological,sohal2020entanglement}, or other methods \cite{kapustin2011topological, fliss2017interface,grosfeld2009nonabelian,lan2015gapped}. 
On the other hand, microscopic approaches are used less often \cite{crepel2019microscopic,
crepel2019model,crepel2019variational,
liang2019parafermions,jaworowski2020model,zhu2020topological,simon2020energetics}. One of the reasons is the required system size (we need to have two well-defined topological orders) combined with reduction of the translational symmetry, which makes the exact diagonalization challenging. 

The interfaces involving non-Abelian FQH states are in general less understood than those of Abelian states only. In addition to general considerations based on topological symmetry breaking  \cite{bais2012modular,bais2009condensate, burnell2018anyon, bais2010topological, burnell2012phase, bais2009theory}, or other methods \cite{lan2015gapped}, particular attention was devoted to the Pfaffian/anti-Pfaffian \cite{barkeshli2015particle,wan2016striped,lian2018theory,mross2018theory, wang2018topological,zhu2020topological,simon2020energetics,hsin2020effective}, Pfaffian/Halperin \cite{yang2017interface,crepel2019variational}, Pfaffian/NASS \cite{grosfeld2009nonabelian} and Pfaffian/Pfaffian \cite{sohal2020entanglement} cases. Few of these calculations were performed on the microscopic level: the Pfaffian/anti-Pfaffian case was studied using DMRG \cite{zhu2020topological,simon2020energetics}, and for the Pfaffian/Halperin case model wavefunctions were created by combining exact MPS matrices obtained from conformal field theory (CFT) \cite{crepel2019variational} (see also \cite{crepel2019microscopic,
crepel2019model} for the same method used for an Abelian interface). Thus, further microscopic studies using various methods are needed to extend our understanding of non-Abelian interfaces and to provide concrete examples of systems embodying the abstract concepts of topological symmetry breaking and related approaches.

Here, we perform a microscopic analysis of a system which, as far as we know, was not studied before – the gapless interface between the non-Abelian Moore-Read state and the Abelian Laughlin state.  While Refs. \cite{crepel2019variational,zhu2020topological,simon2020energetics}  considered continuum systems, we investigate the lattice case, more suited for settings such as optical lattices. This work is also -- up to our knowledge -- the first microscopic study of bulk non-Abelian anyons in the presence of FQH interfaces, as Refs. \cite{crepel2019variational,zhu2020topological,simon2020energetics} focused on the ground state and interface modes. Therefore, it complements the topological symmetry breaking analyses \cite{bais2009condensate, bais2009theory, bais2010topological, burnell2012phase,bais2012modular, burnell2018anyon} which describe anyons and interfaces on a general and abstract level. 

Our results are obtained by using conformal field theory to construct model wavefunctions, and then evaluating their properties using Monte Carlo methods, as we did before for the Abelian Laughlin/Laughlin interface \cite{jaworowski2020model}. This approach is similar to the one by Regnault et. al \cite{crepel2019microscopic,
crepel2019model, crepel2019variational}, but different from it -- we patch together the CFT operators directly, without an additional step of creating MPS, which gives us more freedom of choosing the geometry of the system (we can consider both a plane and a cylinder, with any shape of the interface).

The article is organized as follows. In Section \ref{sec:ground} we construct the ground state wavefunction and investigate its properties, such as particle density profile, correlation function and entanglement entropy for various positions of the cut. Next, in Section \ref{sec:anyons}, we study the anyonic excitations. We determine their charge, statistics and density profile and show that the non-Abelian anyons cannot cross the interface, as their statistics become undefined. In Section \ref{sec:islands} we describe a system with more than one Laughlin island within the MR plane, arguing that it exhibits topological degeneracy. Section \ref{sec:conclusions} summarizes the conclusions of our study.

\section{The ground state}\label{sec:ground}

\subsection{The construction of the wavefunction}
Our approach follows the CFT construction for lattice quantum Hall wavefunctions from Refs.\ \cite{tu2014lattice,glasser2015exact,jaworowski2020model}, which is based on the idea devised by Moore and Read for continuum quantum Hall wavefunctions \cite{moore1991nonabelions}. We consider a system of $N$ sites on a plane, each one at a complex position $z_j=x_j+iy_j$. The sites can be occupied by at most one particle, with $n_j\in\{0,1\}$ denoting the occupation number of site $j$. That is, each site hosts a fermionic or hardcore bosonic degree of freedom (a system can contain sites of both types). We set the charge of each particle to unity. In contrast to the continuum case, we do not assume a constant magnetic field, but rather set it to zero everywhere except at the positions of the lattice sites. That is, we attach an infinitely thin solenoid, containing $\eta_i\in \mathbb{R}^+$ flux quanta, to each site. The total number of flux quanta is $N_{\phi}=\sum_i\eta_i$ and can be different from $N$, which means that we can define two kinds of filling factors. In the simplest case of a single MR or Laughlin state, the particle number $M$ is conserved, so we can define a ``lattice filling'' $\nu_{\mathrm{lat}}=M/N$ and a  ``topological filling'' $\nu=M/N_{\phi}=1/q$, $q\in \mathbb{N}^{+}$. The former describes the degree of discretization (the lower $\nu_{\mathrm{lat}}$, the denser are the lattice points, i.e.\ the closer we are to the continuum), while the latter tells us which state we discretize (i.e.\ it is equal to the Landau level filling factor of the corresponding continuum state).

Any wavefunction in our system can be written in the occupation number basis,
\begin{equation}
\ket{\Psi}=\frac{1}{C}\sum_{\mathbf{n}}\Psi(\mathbf{n})\ket{\mathbf{n}}
\label{eq:wfn}
\end{equation}
where $\mathbf{n}$ is a vector of occupation numbers of all sites, $\ket{\mathbf{n}}$ is a corresponding basis state (to define it unambiguously we fix the order of fermionic creation operators in the definition of $\ket{\mathbf{n}}$ to be the same as the order of site indices),  $\Psi(\mathbf{n})$ are the unnormalized wavefunction coefficients and $C$ is the normalization constant (which we fix to be real without loss of generality). 

To construct a model CFT wavefunction for a single lattice quantum Hall state, an operator $V_{i}(z_i, n_i)$, containing the vertex operator of a certain conformal field theory (depending on which kind of state we want to create), is assigned to each site. The wavefunction is then given by the correlator of these operators for all sites,
\begin{equation}
\Psi(\mathbf{n})=\braket{0|\prod_{i} V_{i}(z_i, n_i) |0}.
\label{eq:correlator}
\end{equation}

This method can be generalized to interfaces \cite{jaworowski2020model}. In general, for two given quantum Hall states, there can be many different types of interfaces. A wavefunction for a particular type can be created by forming a correlator of the form \eqref{eq:correlator}, but made from the operators belonging to two different CFTs. Such a quantity is well-defined when the two CFTs can be embedded in a third one, which puts a restriction on the states for which this method can be applied. We note that in Refs.\ \cite{crepel2019microscopic,crepel2019model,crepel2019variational}, model wavefunctions for interfaces (Laughlin/Halperin, Pfaffian/Halperin) in continuous systems were created by patching together infinite-dimensional matrix product states, derived from conformal field theory, representing the two different fractional quantum Hall states. Our approach is similar, but it uses the CFT operators directly, without the need of a matrix-product-state representation.

So far, we employed this method only for Abelian Laughlin states \cite{jaworowski2020model}. Here, we use it to study an interface between a bosonic Laughlin state and a non-Abelian fermionic Moore-Read state, both at topological filling $\nu=1/2$. They are described by $U(1)_2$ and $U(1)_2\times \mathrm{Ising}$ CFTs, respectively. The embedding condition is satisfied, as the $U(1)_2$ part is the same for both states.

We assume that the system consists of two parts. The left one, which consists of the first $N_L$ sites, is described by the MR state. In the right one, consisting of the next $N_R$ sites up to $N=N_L+N_R$, the particles are in the Laughlin state. The number of flux quanta per site is set to a constant within each part, $\eta_i=\eta_L$ for $i\leq N_L$ and $\eta_i=\eta_R$ for $i> N_L$, but it can differ between the parts, i.e.\ we can have $\eta_L\neq \eta_R$. We note that in general the two parts of the system can be of any shape and can be split into disconnected regions, but we will use the $L$ and $R$ labels for simplicity, as this is the geometry that we will study numerically in this work. 

\begin{figure}
\includegraphics[width=0.5\textwidth]{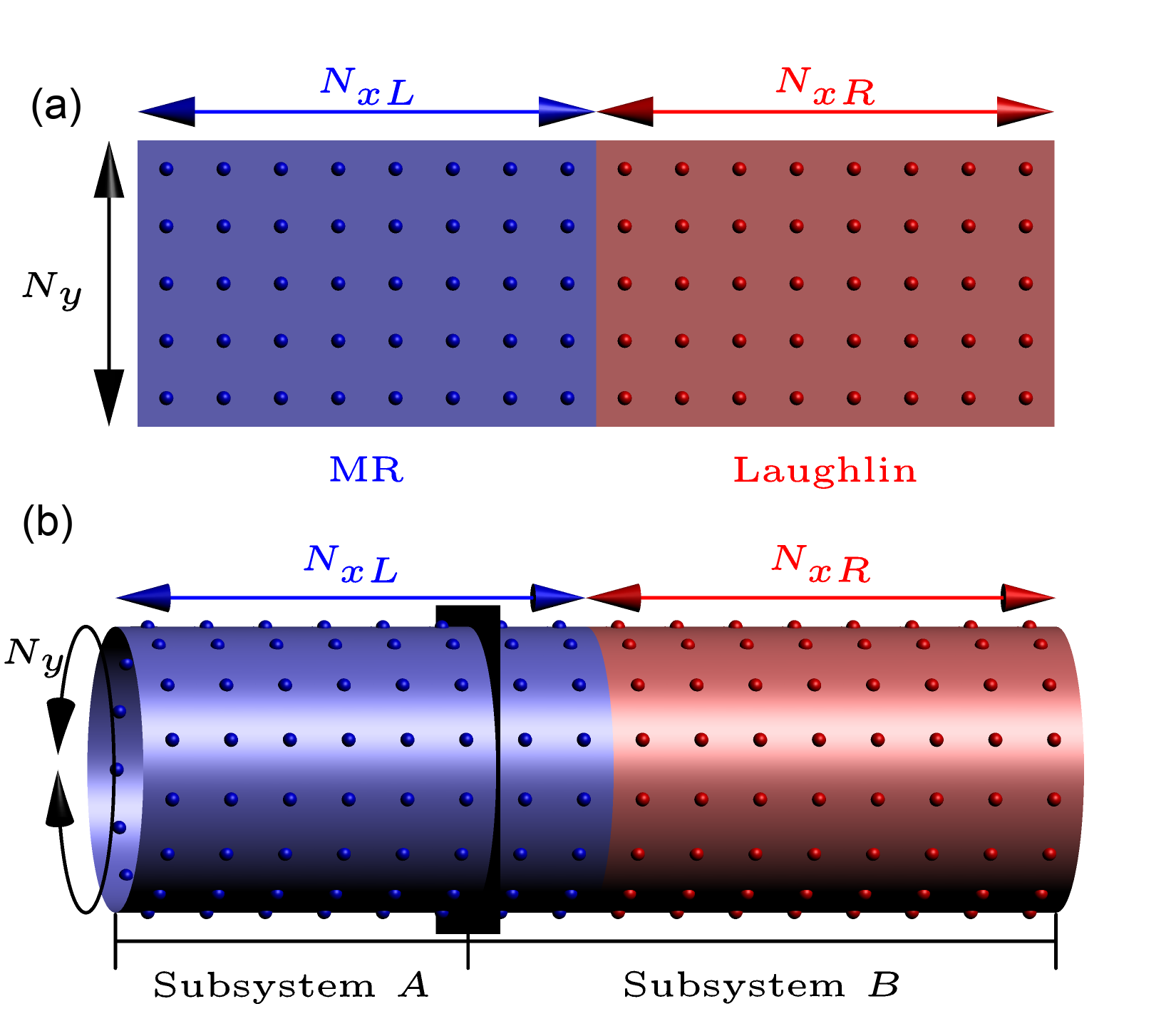}
\caption{The geometry of the considered systems: (a) an example planar system, (b) an example cylindrical system. The vertical plane in (b) shows the division of the system used when computing the entanglement entropy.}
\label{fig:systemview}
\end{figure}

More specifically, the planar systems considered in this work consist of sites arranged in a square lattice of size $(N_{xL}+N_{xR})\times N_y$. The interface is parallel to the $y$ direction, as shown in Fig.\ \ref{fig:systemview} (a). Without loss of generality, we set the lattice constant to unity and the position of the interface to $x=0$.

The operators describing the sites are given by
\begin{equation}
V_{i}(z_i, n_i)=\begin{cases}
V_{\mathrm{Ising}, i}(z_i, n_i) V_{\mathrm{Laughlin}, i}(z_i, n_i)
    & \mathrm{ for }~i\leq N_L \\
V_{\mathrm{Laughlin}, i}(z_i, n_i)   & \mathrm{ for }~i>N_L
\end{cases}.
\label{eq:vertex}
\end{equation}
Here, the $V_{\mathrm{Laughlin}, i}(z_i, n_i)$ and $V_{\mathrm{Ising}, i}(z_i, n_i)$ are the Laughlin-like and Ising-like parts of the operator, respectively. Here
\begin{multline}
V_{\mathrm{Laughlin}, i}=\\=\begin{cases}
e^{i \pi (j-1)\eta_L n_i} \normord{e^{\frac{qn_i-\eta_L}{\sqrt{q}}\phi(z_i)}}    & \mathrm{ for }~i\leq N_L \\
e^{i \pi (j-1)\eta_R n_i} \normord{e^{\frac{qn_i-\eta_R}{\sqrt{q}}\phi(z_i)}}    & \mathrm{ for }~i>N_L 
\end{cases},
\label{eq:vertex_laughlin}
\end{multline}
with $\phi(z_i)$ being a free chiral bosonic field and $q=2$ in our case, ensuring the topological filling $\nu=1/2$ on both sides. We note that the same expressions can be used in the case $q=1$ (an interface between a bosonic MR state and a fermionic integer quantum Hall state, both at $\nu=1$), but then double occupancy of the $L$ sites has to be allowed \cite{glasser2015exact}. In this work we restrict the study to the $q=2$ case. 

The Laughlin-like part is the same for all sites, except from the different values of $\eta_i$ on the two sides. The Ising part is
\begin{equation}
V_{\mathrm{Ising}, i}(z_i, n_i)=\psi(z_i)^{n_i}, 
\end{equation}
where $\psi(z_i)$ is a chiral Majorana field. In contrast to the Laughlin factor, the Ising one is assigned only to the $L$ sites.

By evaluating the correlator \eqref{eq:correlator}, we obtain the unnormalized wavefunction coefficients
\begin{multline}
\Psi(\mathbf{n})=\delta_\mathbf{n}\mathrm{Pf}\left(\frac{1}{z_i'-z_j'}\right) \prod_{i<j}(z_i-z_j)^{qn_in_j} \times \\ \times
\prod_{j=1}^{N_L}\prod_{i(\neq j)}(z_i-z_j)^{-n_i\eta_L}\prod_{j=N_L+1}^{N}\prod_{i(\neq j)}(z_i-z_j)^{-n_i\eta_R},
\label{eq:pfafflin}
\end{multline}
where $i(\neq j)$ means that we sum over all possible values of $i$ (from 1 to $N$) except $i=j$, $z_i', z_j'$ denote coordinates of filled $L$ sites, and 
\begin{equation}
\delta_{\mathbf{n}}=\delta(qM-N_L\eta_L-N_R\eta_R)
\label{eq:cn}
\end{equation}
is the charge neutrality condition ($M=\sum_in_i$ is the total number of particles in the entire system). The charge of particles is compensated by a background charge $-\eta_i/q$ assigned to every site $i$.

We note that, in contrast to the Laughlin/Laughlin interface \cite{jaworowski2020model}, in the Pfaffian/Laughlin case the charge neutrality condition \eqref{eq:cn} enforces the conservation of the total particle number. The numbers of particles $M_I$, $I=L,R$, on each side of the interface can nevertheless fluctuate - provided the same number of particles is annihilated on one side and created on the other. This means that the number of bosons and fermions in the system is not conserved. However, because the Pfaffian in \eqref{eq:pfafflin} is nonzero only if the number of particles in the $L$ part is even, the fermionic parity is conserved, i.e.\ the particles can be created and annihiliated only in pairs.

So far, we have worked on a plane. However, the numerical investigation of certain properties of our wavefunction (such as the entanglement entropy scaling) is easier for a cylinder. We start with an $L_x \times L_y$ rectangle on the complex plane, within which we put the sites at the positions $W_j=\tilde{x}_j+i\tilde{y}_j$. To impose periodic boundary conditions in the $y$ (i.e.\ imaginary) direction, we consider the following mapping
\begin{equation}
z_j=\exp(2\pi W_j / L_y).
\end{equation}
The resulting $z_j$'s are then substituted to \eqref{eq:pfafflin}. We will consider systems on a square lattice, of size $(N_{xL}+N_{xR})\times N_y$ (here $L_y=N_y$), with the interface being parallel to the periodic direction (see Fig.\ \ref{fig:systemview}).

We require that our wavefunction is scale-invariant on a plane and inversion-invariant on a cylinder. The scale invariance is typical for CFT wavefunctions, and it means that the only length scale in the system is the ratio of the magnetic length to the lattice constant, which is set by $\eta_L,\eta_R$. This is the case also in the Hofstadter problem, which is a natural setting for the lattice quantum Hall states, therefore it seems desirable that our wavefunctions exhibit this property. The inversion invariance on the cylinder is also expected, as the physics should stay the same when the Moore-Read and Laughlin parts change places. These requirements are enforced by demanding that the wavefunction does not change  (except from a multiplication by a constant) when subjected to the transformations $z_i\rightarrow cz_i$, $c\in\mathbb{C}$ and $z_i\rightarrow 1/z_i$. It turns out that to fulfill both conditions, one has to set $\eta_L=3/2$, $\eta_R=1$. We enforce this condition throughout this work.

We note that at these values of $\eta$, the possible system sizes can be divided into four classes, corresponding to  $(N_L~\mathrm{mod}~8) = 0,2,4,6$, differing by the way the charge is distributed among $L$ and $R$ parts within each configuration $\ket{\mathbf{n}}$. Each such configuration has well-defined $M_L$ and thus it corresponds to the $L$ charge $M_L-\frac{3}{4}N_L$, as a background charge $-\eta/q=-\frac{3}{4}$ is associated with every $L$ site, and each particle has unit charge.  The charge of the $L$ part is not well-defined for the entire $\ket{\Psi}$ state, as it is a linear combination of different configurations with different values of $M_L$, because pairs of particles can be transferred across the interface. However, the $L$ charge modulo 2, $\Delta Q=\left(\left(M_L-\frac{3}{4}N_L\right)~\mathrm{mod}~2 \right)$ is well-defined, and equal to $\Delta Q=0, 0.5, 1, 1.5$ for $(N_L~\mathrm{mod}~8)= 0,2,4,6$. In other words, only in the first class both parts of the system can be charge-neutral by themselves if the system is cut through the interface. In the third class, the charge neutrality is achieved if a unit charge (e.g. in the form of a quasiparticle) is introduced in each part. In the two remaining classes, such charge would have to be fractional.

The coupling between the $L$ and $R$ parts can be controlled by adjusting the distance between them. If we separate them infinitely far apart from each other, then for the $(N_L~\mathrm{mod}~8)=0$ case (i.e. when charge neutrality on each side can be satisfied separately), the wavefunction becomes a tensor product of MR and Laughlin states (see Appendix \ref{app:copuling}). In this work, we consider the situation where $L$ and $R$ are separated by one lattice constant, i.e. the entire system is the perfect square lattice as in Fig.\ \ref{fig:systemview}.

Finally, let us remark that for some values of $\eta$, it is possible to derive a parent Hamiltonian for single Moore-Read \cite{glasser2015exact,manna2018nonabelian} or Laughlin \cite{glasser2016lattice} lattice quantum Hall states. However, it is not straightforward to extend these calculations to the case when the system is described by two different CFTs. Another way to connect our wavefunction to a Hamiltonian is to find a short-range Hamiltonian, whose ground state is approximated by our wavefunction, following the approach from Ref. \cite{glasser2015exact}: diagonalizing the Hamiltonian numerically and optimizing its coefficients to maximize the overlap between the ground state and our wavefunction. However, this would require extensive exact-diagonalization or DMRG calculations, and we leave it for future works.

Often the wavefunction itself, which may, but does not have to, be related to a Hamiltonian, can provide insights into the inner structure of given topological order. For example, the Laughlin wavefunction revealed the physical mechanism of FQHE and the nature of the fractionalized excitations \cite{laughlin}. The Kalmeyer-Laughlin wavefunction provided a vital example of a chiral spin liquid \cite{kalmeyer1987equivalence, kalmeyer1989theory}. The concepts of Haldane hierarchy \cite{haldane1983fractional} or composite fermions \cite{jain1989composite} were also embodied in wavefunctions. Speaking of the interfaces, Regnault et al. used a model wavefuncion to study the nature of the interface modes \cite{crepel2019model,crepel2019microscopic,crepel2019variational}, while we employed our model state to determine the properties of bulk anyons in presence of the interface \cite{jaworowski2020model}. Therefore, we believe that the study of the wavefunction itself is important and thus, in this work we focus solely on $\ket{\Psi}$.

\begin{figure}
\includegraphics[width=0.5\textwidth]{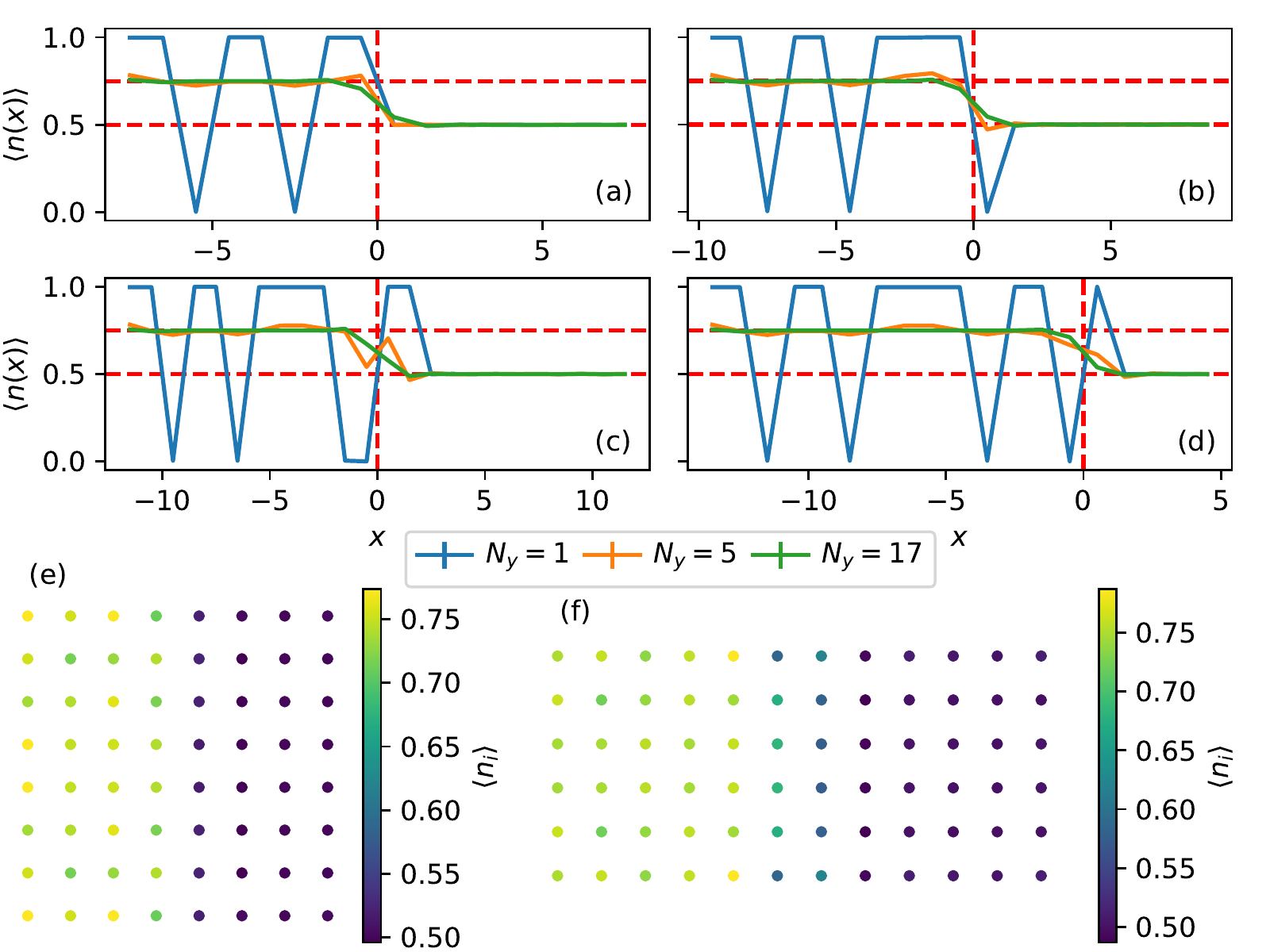}
\caption{The particle density plots for the ground state of systems with an interface. (a)-(d) The density as a function of $x$ for example cylindrical systems from four groups, (a), (b), (c), (d), corresponding to $(N_L~\mathrm{mod}~8)=0,2,4,6$. The color denotes $N_y$. The $N_{xR}$ size of the systems within a group is not necessarily the same. (e), (f) Example plots of particle density for planar systems of size $(4+4)\times 8$ and $(6+6)\times 6$, respectively. 
}
\label{fig:DensityGS}
\end{figure}

\subsection{Numerical results -- particle density}\label{ssec:GSParticleDensity}
Once we have the wavefunction \eqref{eq:pfafflin}, we can study its properties numerically using Monte Carlo methods. In particular, it is straightforward to obtain the average particle density $\langle n_i \rangle$. On a cylinder, the density is constant in the $y$ direction, so we define the density as a function of $x$
\begin{equation}
\langle n(x) \rangle=\frac{1}{N_y}\sum_i \langle n_i \rangle \delta(x-x_i)
\end{equation}

We investigate this quantity for a number of systems with different sizes, some of which are shown in Fig.\ \ref{fig:DensityGS} (a)-(d). When the cylinder is thin, the states display large oscillations in density within the $L$ part. In particular, at $N_y=1$ (blue curves in \ref{fig:DensityGS} (a)-(d)), we have either $\langle n_i \rangle =0$ or $\langle n_i \rangle =1$, with no fractional values, reminiscent of the thin torus limit of the continuum FQH states. As the cylinder gets wider, the density in the $L$ bulk becomes close to $3/4$, as expected for a $\eta=3/2$, $\nu=1/2$ MR state (see the orange and green curves in Fig.\ \ref{fig:DensityGS} (a)-(d)). Apart from very thin cylinders, the density inhomogenities exist mostly near the edges and the interface and get smaller as $N_y$ increases. In the $R$ part, independently of $N_y$, the density is close to $1/2$ everywhere except from the vicinity of the interface. This value is expected for a $\eta=1$, $\nu=1/2$ Laughlin state.

Near the interface, the density has to drop from  3/4 to 1/2. How this happens exactly depends on the size of the system. As can be seen in Fig.\ \ref{fig:DensityGS} (a)-(d), for thin enough cylinders (e.g. $N_y=5$), the four $(N_L~\mathrm{mod}~8)$ groups display four qualitatively different patterns of particle density. This is most striking when comparing $(N_L~\mathrm{mod}~8)=4$, which has an additional ``step'' (i.e.\ a local maximum of density near the interface in part $R$) to $(N_L~\mathrm{mod}~8)=0$, where such a feature is absent. For wider cylinders (e.g. $N_y=17$), the density profiles in all four groups become similar.

In the case of planar systems, the density patterns are more complicated, as the translational invariance is lost, and the density inhomogenities exist near the interface and all three edges of the $L$ part. Two examples are shown in Fig.\ \ref{fig:DensityGS} (e),(f).

\begin{figure}
\includegraphics[width=0.5\textwidth]{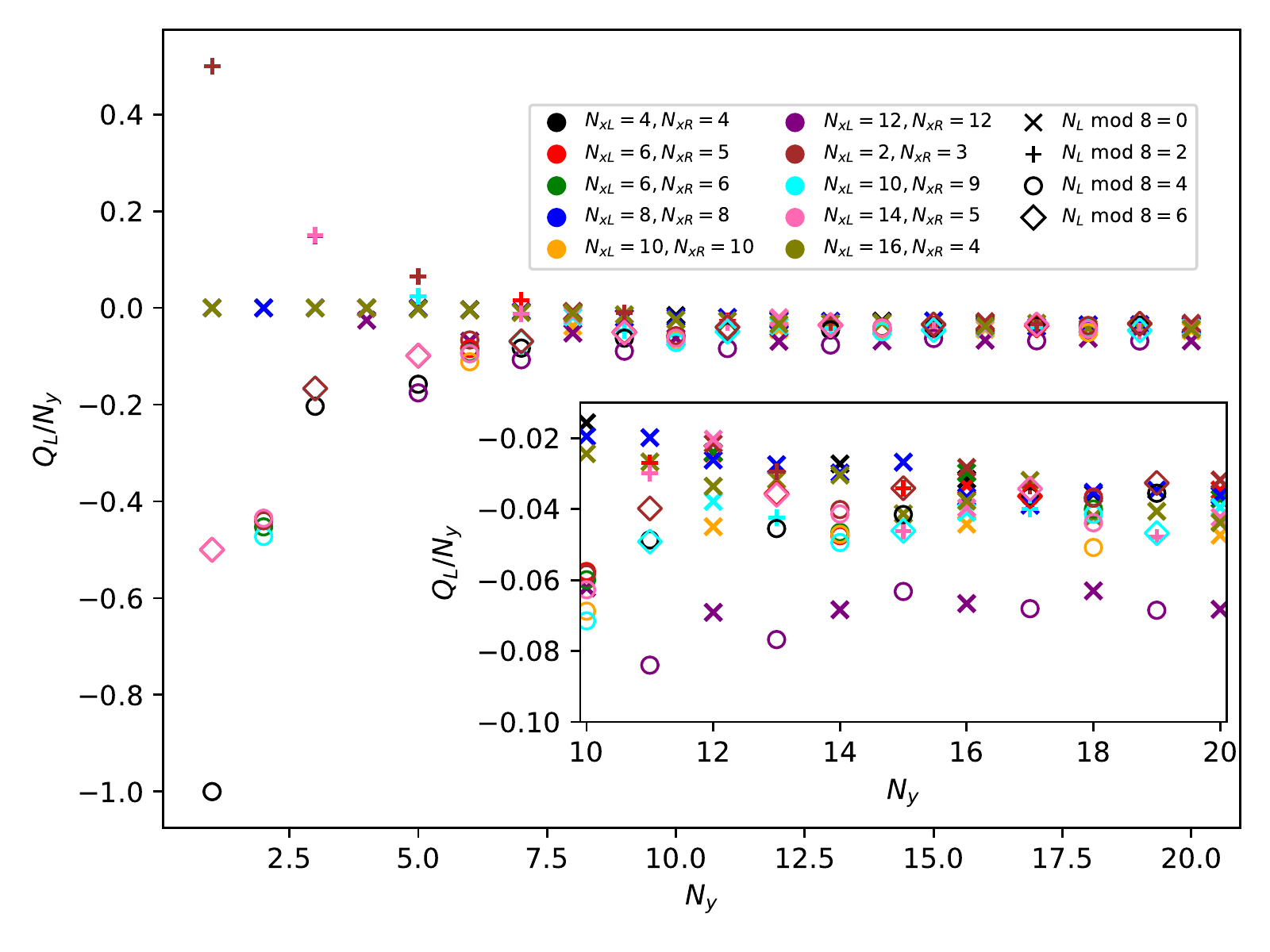}
%\caption{The total charge $Q_L$ in part $L$ as a function of total number of sites $N$. The different colors correspond to different system sizes in the $x$ direction. The marker shapes denote the values of $N_L~\mathrm{mod}~8$.
\caption{The total charge $Q_L$ in part $L$ per unit of interface length, as a function of interface length $N_y$. The different colors correspond to different system sizes in the $x$ direction. The marker shapes denote the values of $N_L~\mathrm{mod}~8$. The inset shows the magnification of the right part of the plot.
}
\label{fig:QLGS}
\end{figure}

The background charge $-\eta_i/q$ changes abruptly from $-3/4$ to $-1/2$ at the interface. On the other hand, we have seen that the particle density changes more smoothly. This means that some excess charge is accumulated near the interface on each side. Let us investigate this more closely for the cylindrical systems we have already studied. We define the excess charge as a function of $x$ as
\begin{equation}
Q(x)=\sum_i (\langle n_i \rangle -\eta_i/q) \delta(x-x_i),
\end{equation}
and a total charge accumulated in part $I$ as 
\begin{equation}
Q_I=\sum_{i\in I} (\langle n_i \rangle -\eta_I/q).
\end{equation}

%Apart from thin cylinders,the excess charge is concentrated mostly near the interface $x=\pm 0.5$, with $Q(-0.5)\approx -Q(0.5)$. Thus, for wide enough cylinders, we have $Q(-0.5)\approx Q_L$. Therefore, let us study the latter quantity as a function of system size. The results are shown in Fig.\ \ref{fig:QLGS}. The different colors correspond to different system sizes in the $x$ direction, while the marker shapes refer to the four classes $N_L~\mathrm{mod}~8$. It can be seen that at $N_y=1$ (the rightmost points from every class) the charge is equal $\Delta Q$. When increasing $N_y$, the charge starts to vary, depending strongly on the system size and shape. For large (and wide) enough systems, $Q_L$ is negative in all four classes. Moreover, there seems to be an upper bound for the charge, decreasing as the total size $N$ of the system increases. For the investigated systems, the $Q_L$ seems to approach the upper bound as $N$ increases. 
%
%The charge $Q(-0.5)$ behaves very similarly to Fig.\ \ref{fig:QLGS} for large cylinders.   Qualitative differences arise in the $N_y=1$ limit, as then $Q(-0.5)=1/4$ (filled sites at $x=-0.5$) or $Q(-0.5)=-3/4$ (empty sites at $x=-0.5$).
%
%The fact that for wide enough cylinders the accumulated charge depends on $N$, not $N_y$ (i.e.\ on both dimensions of the cylinder, taking into account both parts, instead of the length of the interface), suggests that our wavefunction does not necessarily have a well-defined thermodynamic limit. That is, 

Apart from thin cylinders, the excess charge is concentrated mostly near the interface, at $x=\pm 0.5$, with $Q(-0.5)\approx -Q(0.5)$. Thus, for wide enough cylinders, we have $Q(-0.5)\approx Q_L$. Therefore, let us study the latter quantity as a function of system size. To be precise, instead of $Q_L$ itself, we investigate the charge in part $L$ per unit of interface length, i.e.\ $Q_L/N_y$. If there is a fixed density pattern in the thermodynamic limit, then this quantity should converge to a fixed value. 

The results are shown in Fig.\ \ref{fig:QLGS}. The different colors correspond to different system sizes in the $x$ direction, while the marker shapes refer to the four classes $(N_L~\mathrm{mod}~8)$. It can be seen that at $N_y=1$ (the leftmost points from every class) the charge modulo 2 equals $\Delta Q$. As we increase $N_y$, $Q_L/N_y$ in all four classes seems to display convergence towards a fixed, negative value of $Q_L/N_y$, lying between $-0.03$ and $-0.07$.

The charge $Q(-0.5)$ behaves very similarly to Fig.\ \ref{fig:QLGS} for wide cylinders.   On the other hand, qualitative differences arise in the thin cylinder limit. At $N_y=1$, there are only two possible values: $Q(-0.5)=1/4$ (filled site at $x=-0.5$) or $Q(-0.5)=-3/4$ (empty site at $x=-0.5$).

A similar charge accumulation at the interface was observed in Ref. \cite{zhu2020topological} for the Pfaffian/anti-Pfaffian case. In this work, the accumulated charge on the two sides of the interface in the thin-torus limit was equal to $\pm 1/4$ (i.e.\ the charge of a non-Abelian quasiparticle). In our case, $Q_L$ is equal to a charge of an Abelian quasiparticle or a particle, depending on $N_L$. %Moreover, in Ref.\ \cite{zhu2020topological} the authors observed that the dipole moment with respect to the interface is constant also outside the thin-torus limit. This dipole moment was of topological origin, as it arose from the difference of the quantized Hall viscosities on both sides of the interface. In contrast, in our work the dipole moment is not constant. This can be seen easily by noting that for large enough $N_y$, $Q(-0.5)/N_y$ is roughly constant, therefore  $Q(-0.5)$ should grow more or less linearly with $N_y$. Since most of the charge is concentrated at $x=\pm 0.5$, this means that the dipole moment must grow with $N_y$. To verify this, we evaluated the dipole moment explicitly, confirming that it is indeed not conserved.

\begin{figure}
\includegraphics[width=0.5\textwidth]{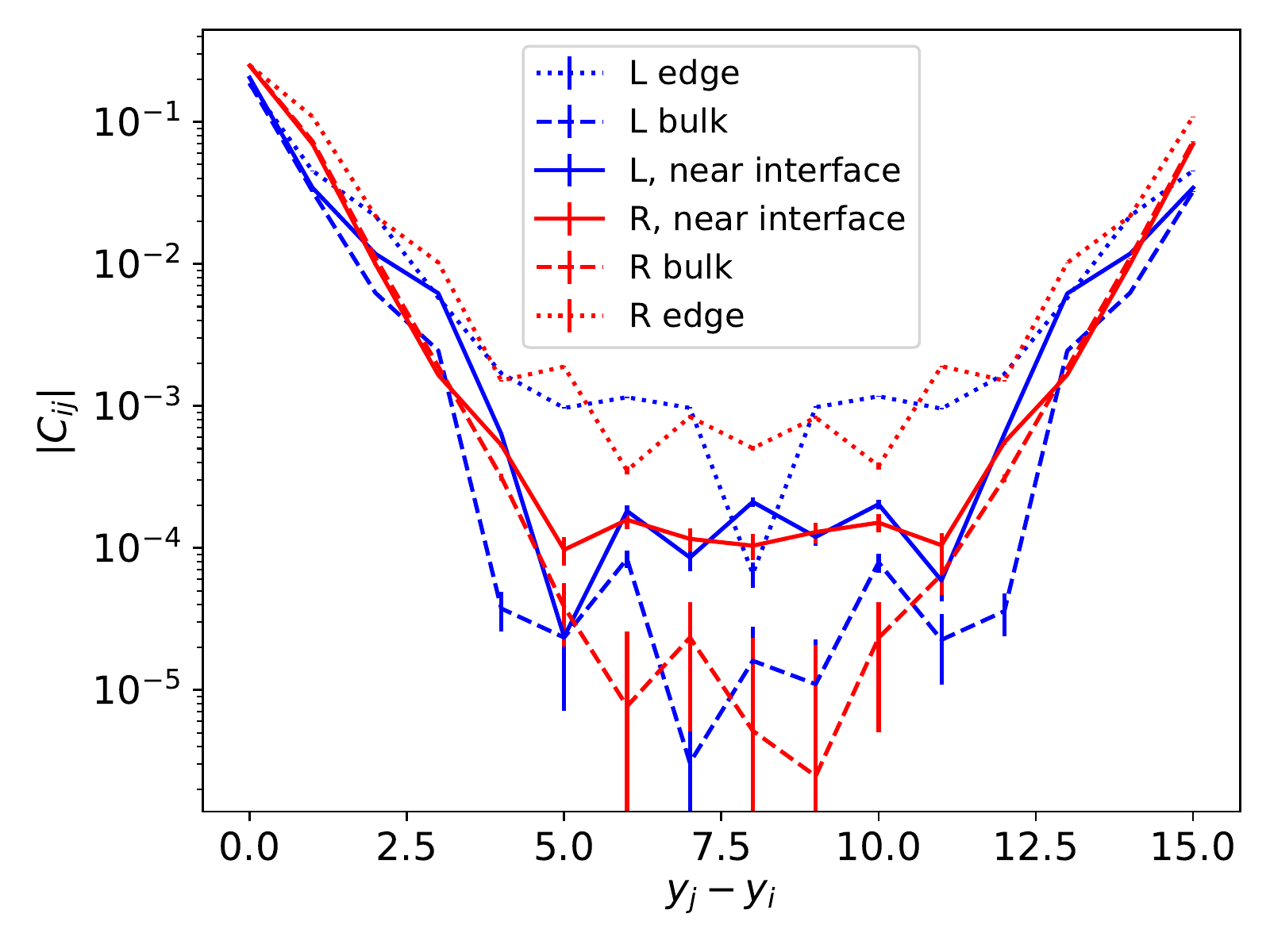}
\caption{The absolute value of correlation function $|C_{ij}|$ for $x_i=x_j$ as a function of $y_j-y_i$ at different locations of $x_i=x_j$: in the bulk of each part ($x=-5,5$, $x=4.5$), at the edges ($x=-10.5$, $x=10.5$) and near the interface ($x=-0.5$, $x=0.5$). The system size is $(10+10)\times 16$}
\label{fig:CorrelationGS}
\end{figure}

\subsection{Numerical results -- correlation function}

We expect that our interface is gapless. This is because the edges of the Laughlin state are described by a chiral Luttinger liquid \cite{wen1992theory}, while the MR state has also a single Majorana fermion edge mode \cite{wen1995topological,milovanovic1996edge}. In the effective interface theories considered so far for various non-Abelian interfaces \cite{yang2017interface,barkeshli2015particle,wan2016striped,lian2018theory,mross2018theory, wang2018topological,sohal2020entanglement}, the Majorana mode can be gapped only when paired with a second Majorana mode. Since there is just a single Majorana mode in the system, we expect that it cannot be gapped.

It is expected that gapped systems generated by short-range interactions have exponentially decaying correlation functions. In our case, we do not have the parent Hamiltonian, so we cannot ensure that our wavefunction indeed can be generated by a short-range interaction. But assuming it is the case, the correlation function would give us some indication on whether the interface is gapped or gapless. 

The correlation function is given by
\begin{equation}
C_{ij}=\langle n_i n_j \rangle - \langle n_i \rangle \langle n_j \rangle,
\end{equation}
and can be easily computed using Monte Carlo. For the ease of presentation, we choose sites with $x_i=x_j$ and investigate $C_{ij}$ as a function of $y_j-y_i$. The results for the different values of $x_i=x_j$ are shown in Fig.\ \ref{fig:CorrelationGS}. 

In the bulk, the correlation function seems to decay roughly exponentially. However, near the edges its decay seems to be slower. Near the interface, the values of the correlation function halfway across the cylinder seem to be located roughly in the middle between the results for the bulk and the edge, still showing the lack of exponential decay. This suggests that the interface is indeed gapless (provided that it is generated by a short-range interaction).

\begin{figure}
\includegraphics[width=0.5\textwidth]{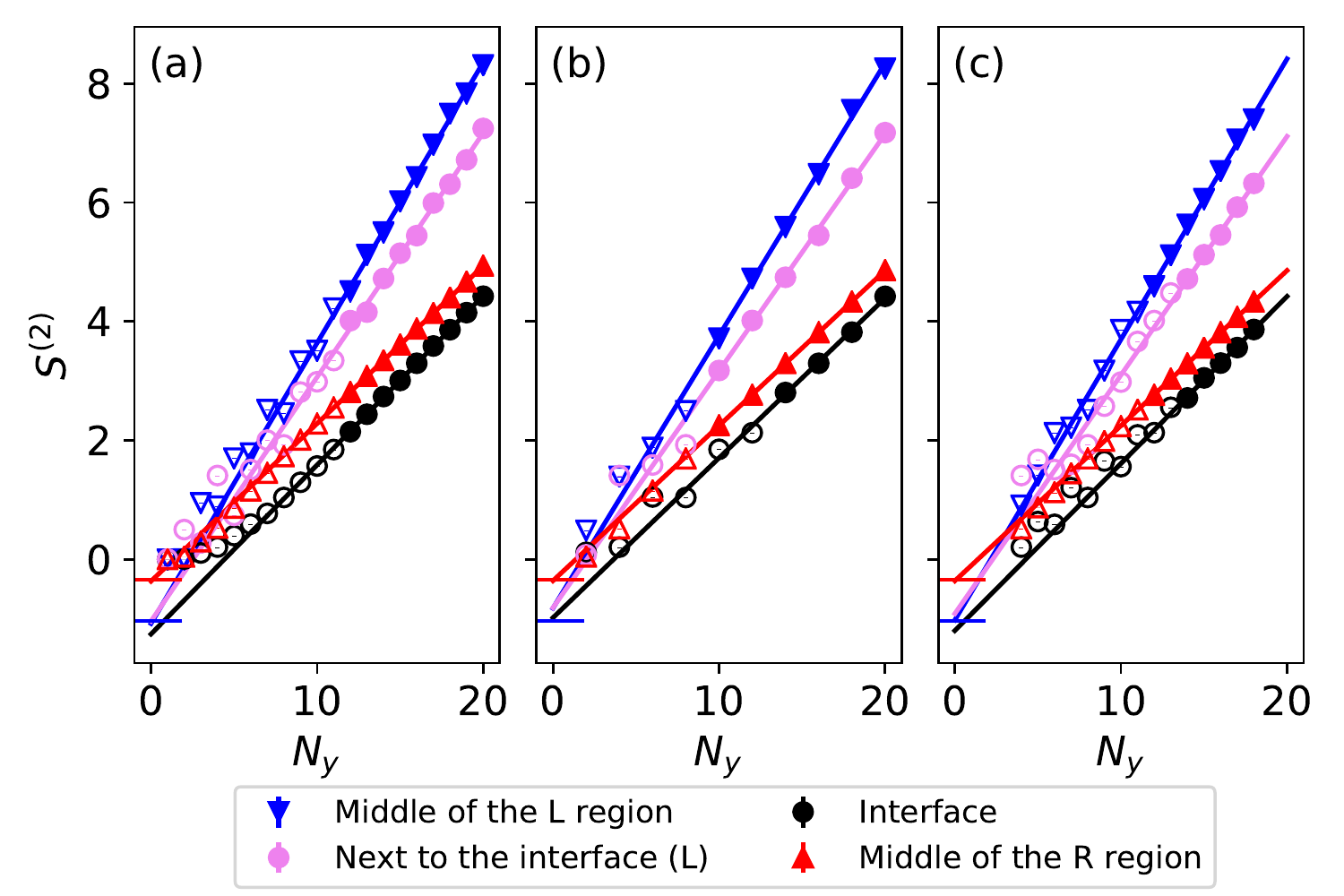}
\caption{The second R\'enyi entropy as a function of cylinder circumference for three series of systems: (a) $(8+8)\times N_y$, (b) $(10+4)\times N_y$ and (c) $(12+4)\times N_y$.}
\label{fig:EntropyGS}
\end{figure}

\subsection{Numerical results -- entanglement entropy}
The topological properties of the interface can manifest themselves in the entanglement entropy when the cut coincides with the interface. While this issue was studied using field theory for Laughlin/Laughlin \cite{cano2015interactions,santos2018symmetry} or Pfaffian/Pfaffian interfaces \cite{sohal2020entanglement}, for the Pfaffian/Laughlin interface, up to our knowledge, there were no predictions how the entropy should scale with the interface length. Thus, we are going to study the entropy numerically, using the Monte Carlo method outlined in \cite{cirac2010infinite, hastings2010measuring} (see also our previous work where we used this method \cite{tu2014lattice,glasser2015exact,jaworowski2020model}). Within this approach, the second R\'enyi entropy can be obtained by sampling two independent copies of the system.

In Fig.\ \ref{fig:EntropyGS}, we show the entanglement entropy scaling for three series of systems, of size: $(8+8)\times N_y$, $(10+4)\times N_y$ and $(12+4)\times N_y$. The cut is parallel to the interface, as shown in Fig.\ \ref{fig:systemview}. We are interested especially in the four positions of the cut: in the bulks of the two sides, precisely at the interface and right next to the interface on the left (i.e.\ $x=-1$).

The scaling in the bulks in Fig.\ \ref{fig:EntropyGS} corresponds to the position $x=-N_{xL}+\lfloor N_{xL}/2\rfloor$ for the $L$ side and $x=\lfloor N_{xR}/2\rfloor$ for the $R$ side, with $\lfloor \rfloor$ denoting the floor function. If the interface wavefunction indeed describes the expected topological orders, then, by applying a linear fit,
\begin{equation}
S^{(2)}(N_y)=AN_y-\gamma
\end{equation}
we should recover the expected value of $\gamma$: $\gamma_L=\ln(8)/2$ and $\gamma_{R}=\ln(2)/2$, corresponding to the topological entanglement entropies of the MR and Laughlin state, respectively. These values are indicated by blue and red ticks, respectively, on the $y$ axes of the plots. The red and blue lines denote the fits. Because for thin cylinders the linear scaling is distorted by finite-size effects, we discard these systems from the calculation. That is, in the fit we include only the data points denoted  by filled symbols. The fits seem to cross the $N_y=0$ line relatively close to the predicted values. For $\gamma_R$, the agreement is good: we obtain
 $0.368 \pm 0.005$, $0.361 \pm 0.009$, $0.36 \pm 0.02$, for $(8+8)\times N_y$, $(10+4)\times N_y$ and $(12+4)\times N_y$, respectively, compared to $\ln(2)/2\approx 0.347$. The uncertainties here are the fit uncertainties, without the inclusion of Monte Carlo uncertainties. This confirms that the $R$ part has a $\nu=1/2$ Laughlin-type topological order.
 
For $\gamma_L$, we obtain $1.08 \pm 0.13$, $0.82 \pm 0.17$, $1.02 \pm 0.15$, respectively, compared to $\ln(8)/2\approx 1.03$. That is, the agreement is worse, and the error bars are much bigger. In addition, the result for part $L$ seems to depend strongly on the position of the cut and on which data points we take into account on the fit. Also, while the fits in Fig.\ \ref{fig:EntropyGS} were performed without the inclusion of MC error bars in the weights, including them makes the result even more dependent on the number of included data points. The detailed analysis is contained in Appendix \ref{app:entropy}. Nevertheless, the fitted values oscillate around the predicted value and are clearly nonzero. Thus, we conclude that the $L$ part is also topologically ordered, and the results are consistent with the Moore-Read topological order, although not indicating it clearly.

What happens with the entropy when the cut coincides with the interface (black markers in Fig.\ \ref{fig:EntropyGS})? For almost all the investigated systems, the entropy at the interface is lower than in the bulks of both sides (excluding some thin cylinders). However, as $N_y$ increases, the interface entropy increases faster than the $R$ bulk entropy, thus we can expect that the former will finally dominate over the latter. For large enough $N_y$, the scaling seems to be linear. Because we do not have compelling theoretical arguments that in this case such a scaling is expected in the thermodynamic limit, we do not rule out the possibility that the perceived linear dependence is in fact nonlinear, and the nonlinearity would show up for larger $N_y$. Nevertheless, assuming that it is linear, we perform the fit. The obtained values are close to $\ln(8)/2$, i.e.\ the topological entanglement entropy of the left part. If this is indeed the case, this is similar to the case of Laughlin states at fillings $1/q_L$, $1/q_R$ such that $q_R=a^2q_L$, $a\in \mathbb{N}^{+}$ \cite{cano2015interactions,santos2018symmetry,jaworowski2020model}. However, the fitted parameters are subjected to the same distortions and uncertainties as $\gamma_R$, thus we cannot conclude that it is indeed the case.

How far does the influence of the interface extends into the $L$ and $R$ parts?  Next to the interface on the right ($x=1$), the entanglement entropy values for large enough $N_y$ are similar to the ones in the bulk $R$ part. However, on the left, the influence of the interface is apparent in the first column of sites next to it ($x=-1$). The values of the entanglement entropy (except from some low-$N_y$ systems) are lower than in the $L$ bulk of the same system (see the violet markers in Fig.\ \ref{fig:EntropyGS}). The fit for large $N_y$  also yields values roughly close to the theoretical value of $\gamma_L$, although again the results are uncertain due to the dependence of the fitted value on the data points included. Thus, we do not rule out the possibility that near the interface there might be some variation of $\gamma$, e.g. a similar increase as in the Laughlin-Laughlin interfaces \cite{jaworowski2020model}.

\section{The systems with anyons}\label{sec:anyons}
Having determined the ground state wavefunctions, we now wonder, what are the properties of the  anyonic excitations above the ground state.
\subsection{The construction of the wavefunction}\label{ssec:AnyonWfn}
The wavefunctions including anyons can be obtained by inserting further operators into the correlator \eqref{eq:correlator}. These operators depend on parameters $w_i$, the complex coordinates of the anyons. The state is given by
\begin{equation}
\ket{\Psi}_\alpha=\frac{1}{C_\alpha}\sum_{\mathbf{n}}\Psi_\alpha(\mathbf{n},\mathbf{w}) \ket{\mathbf{n}}.
\label{eq:AnyonWfn}
\end{equation}
There are three differences between \eqref{eq:AnyonWfn} and \eqref{eq:wfn}. First, now the wavefunction coefficients, as well as the normalization constant, depend on the external parameters, the anyon coordinates $\mathbf{w}$. Secondly, there can be more than one degenerate state, hence we introduced the index $\alpha$. Third, while for the ground state the fermion parity conservation was enforced by the Pfaffian factor, in the presence of anyons the correlators are nonzero both for even and odd $M_L$. Thus, in general, we can construct a wavefunction which does not conserve fermion parity. However, we expect that it would be unphysical and hence it cannot be generated by a local Hamiltonian. Moreover, as we will see later, it would generate problems with boundary conditions for anyons. To restore the fermion parity conservation, we assume that the interaction generating our wavefunction allows to exchange particles through the interface only in pairs. Then, the Hilbert space divides into two disconnected parts, with even and odd $M_L$. We focus on the case of even $M_L$.

While the particle coordinates are restricted to the lattice sites, the anyon coordinates can be located anywhere on the plane/cylinder. In such a way, we will be able to move them smoothly, which will be important when evaluating their statistics.

We study two classes of anyons of the Moore-Read state. The basic non-Abelian excitations are constructed using the following operator \cite{manna2018nonabelian,manna2019quasielectrons}
\begin{equation}
V_{\mathrm{NA},k}(w_k)=\sigma(w_k)\normord{\exp \left( \frac{p_i}{\sqrt{q}}\phi(w_k)  \right)},
\label{eq:vertex_na}
\end{equation}
where $\sigma(w_k)$ is the holomorphic spin operator of the chiral Ising CFT, and $p_k=1/2$, $p_k=-1/2$, correspond to a quasihole and a quasielectron, respectively. We note that the latter are difficult to construct in the continuum \cite{patra2020continuum}, whereas for the lattice their construction is simple -- it requires only flipping the sign of the $p_k$.

The other group of excitations consists of Laughlin-like Abelian anyons, described by the operator \cite{nielsen2015anyon}
\begin{equation}
V_{\mathrm{A},k}(w_k)=\normord{\exp \left( \frac{p_i}{\sqrt{q}}\phi(w_k)  \right)},
\end{equation}
where $p_k$ is now integer. This operator describes also the excitations of the Laughlin state. Thus, these anyons are valid topological excitations of the entire system. In contrast, the ones generated by the operator \eqref{eq:vertex_na} are valid topological excitations only within the $L$ part. Nevertheless, technically we can also attempt to put such an anyon in the $R$ part and see what happens.

We will refer to these two groups as ``Abelian'' and ``non-Abelian'' for brevity, although the reader should bear in mind that the anyonic content of the Moore-Read state is richer than the considered cases. We denote the numbers of non-Abelian and Abelian anyons as $R_{\mathrm{NA}}$ and $R_{\mathrm{A}}$, and their total number as $R=R_{\mathrm{NA}}+R_{\mathrm{A}}$. For convenience, we will also assume that the anyons are indexed in such a way that the first $R_{\mathrm{NA}}$ are non-Abelian, and the rest are Abelian. 

The wavefunction coefficients for even $M_L$ are now given by the following correlator
\begin{multline}
\Psi_\alpha( \mathbf{w},\mathbf{n})=\\=\braket{0|\prod_{i=1}^{R_{\mathrm{NA}}} V_{\mathrm{NA},i}(w_i) \prod_{i=R_{\mathrm{NA}}+1}^{R} V_{\mathrm{A},i}(w_i)  \prod_{i=1}^{N} V_{i}(z_i, n_i) |0}{}_\alpha=\\=I_\alpha( \mathbf{w},\mathbf{n})J( \mathbf{w},\mathbf{n}),
\label{eq:AnyonWfnCoeffs}
\end{multline}
where the index $\alpha$ means that we take only the conformal block where the Ising fields fuse to $\alpha$, and $I_\alpha$ and $J$ are the Ising and Jastrow parts of the wavefunction, respectively.  The latter is given by
\begin{multline}
J( \mathbf{w},\mathbf{n})=\braket{0|\prod_{i=1}^{R} \normord{\exp \left( \frac{p_i}{\sqrt{q}}\phi(z_i)  \right)} \prod_{i=1}^{N} V_{i}(z_i, n_i) |0}=\\
=\delta_{\mathbf{n}}\prod_{i<j}(w_i-w_j)^{p_ip_j/q}  \prod_{i,j}(w_i-z_j)^{p_in_j} \prod_{i,j}(w_i-z_j)^{-p_i\eta_j/q}\times \\ \times
 \prod_{i<j}(z_i-z_j)^{qn_in_j} \prod_{i\neq j}(z_i-z_j)^{-n_i\eta_j},
\end{multline}
where the charge neutrality is enforced by
\begin{equation}
\delta_{\mathbf{n}}=\delta(qM+\sum_{i}p_i-\eta_LN_L-\eta_R N_R).
\label{eq:AnyonCN}
\end{equation}
The Ising part depends on the number of non-Abelian anyons. If there are none, then there is just one conformal block, and we have
\begin{equation}
I( \mathbf{w},\mathbf{n})=\mathrm{Pf}\left(\frac{1}{z_i'-z_j'} \right).
\end{equation}
If $R_{\mathrm{NA}}=2$, then there is also one conformal block, with
\begin{equation}
I( \mathbf{w},\mathbf{n})=
2^{-M_L/2}(w_1-w_2)^{-1/8}\prod_{i=1}^{2} \prod_{j=1}^{N_L}(w_i-z_j)^{-n_j/2}\mathrm{Pf}\mathbf{A},
\label{eq:ising_2qh}
\end{equation}
where \begin{equation}
A_{ij}=\frac{(z'_i-w_1)(z'_j-w_2)+(z'_i-w_2)(z'_j-w_1)}{z_i'-z_j'}.
\end{equation}
If $R_{\mathrm{NA}}=4$, then there are two conformal blocks, corresponding to the situation where the pairs of $\sigma$ fields fuse to $\alpha=I$ or $\alpha=\psi$. The Ising part is given by
\begin{multline}
I_{\alpha}( \mathbf{w},\mathbf{n})=2^{-(M_L+1)/2}(w_1-w_2)^{-1/8}(w_3-w_4)^{-1/8}\times \\ \times
\prod_{i=1}^{4}\prod_{j}(w_i-z_j)^{-n_i/2} \left((1-x)^{1/4} +\frac{(-1)^{m_\alpha}}{(1-x)^{1/4}}\right)^{-1/2} \times \\ \times
\left( (1-x)^{\frac{1}{4}}\Phi_{(13)(24)} + (-1)^{m_\alpha}(1-x)^{-\frac{1}{4}}\Phi_{(14)(23)} \right)
\label{eq:ising_4qh}
\end{multline}
where $m_\alpha=0,1$ for  $\alpha=I,\psi$, respectively,
\begin{equation}
x=\frac{(w_1-w_2)(w_3-w_4)}{(w_1-w_4)(w_3-w_2)},
\end{equation}
and
\begin{multline}
\Phi_{(kl)(mn)}=\\
=\mathrm{Pf} \left( 
\frac{(w_{k}-z'_i )(w_{l}-z'_i )(w_{m}-z'_j )(w_{n}-z'_j )+(i\leftrightarrow j)}{z_i'-z_j'}
\right)
\end{multline}

\begin{figure}
\includegraphics[width=0.5\textwidth]{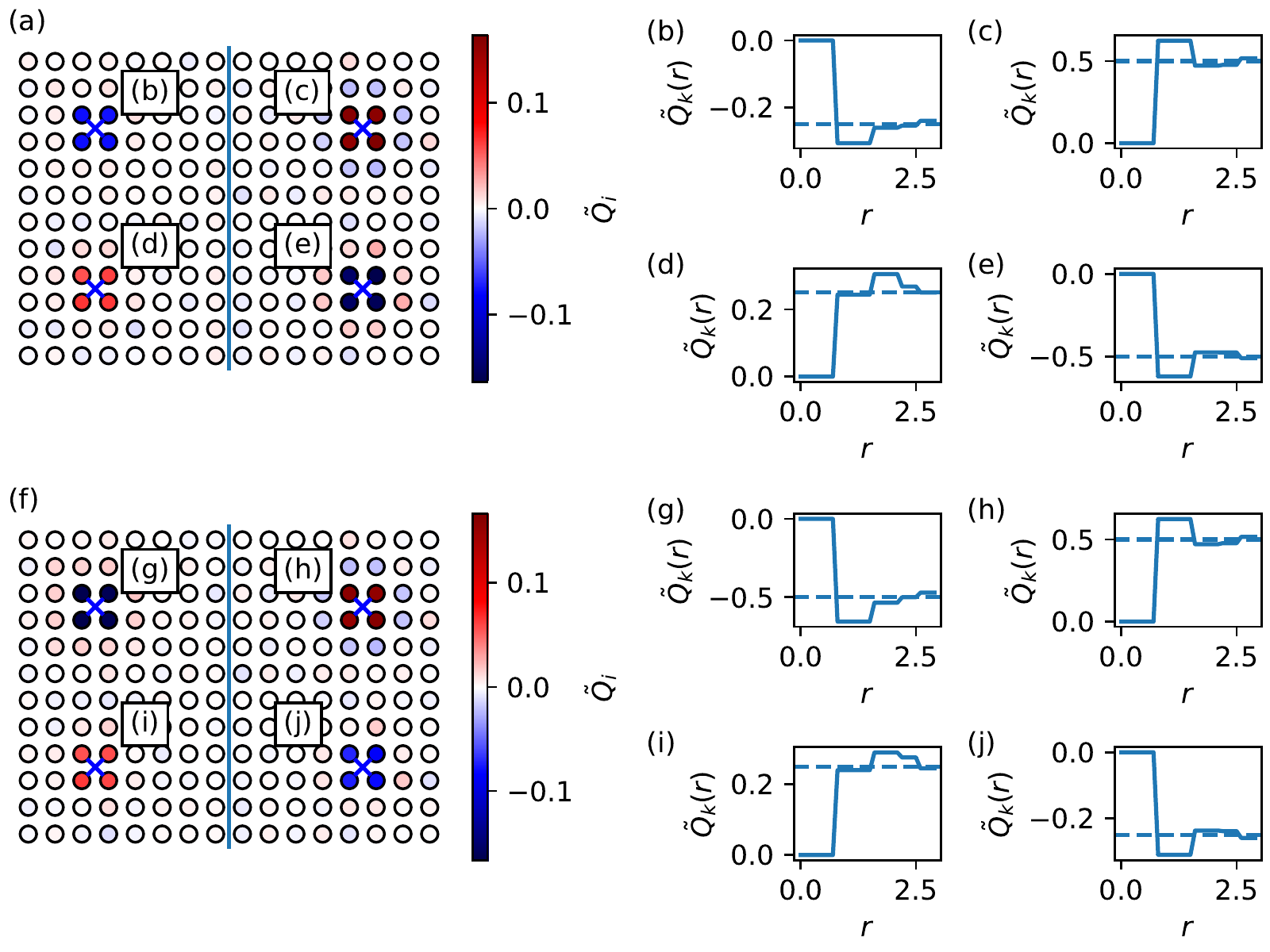}
\caption{(a), (f): the excess charge density in the presence of anyons. (b)-(e) and (g)-(j): the radial excess charge profiles for the anyons denoted by the respective subplot label in (a), (f). The system is a cylinder of size $(8+8)\times 12$ and contain four anyons: the non-Abelian quasihole with $p_k=0.5$ ((b), (j)), the non-Abelian quasielectron with $p_k=-0.5$ ((c),(h)), the Abelian quasielectron with $p_k=-1$ ((d), (i)) and the Abelian quasihole with $p_k=1$ ((e),(g)). The plots (a)-(e) correspond to the case when the non-Abelian anyons are both located in the $L$ part, while in (f)-(j) one of them is located in the Laughlin part. The ``$\times$'' symbols in (a), (f) denote the anyon coordinates. In (b)-(e) and (g)-(j), the horizontal dashed lines denote the expected charges $-p_k/q$.}
\label{fig:AnyonDensityFigure}
\end{figure}

\subsection{Anyon charge and density distribution}\label{ssec:AnyonDensity}
Let us now verify that the anyons are well screened and that their charges agree with the theoretical prediction $-p_k/q$. We define the excess charge at site $i$ in the presence of anyons as the difference 
\begin{equation}
\tilde{Q}_i=\langle n_i \rangle_{\mathrm{an}}-\langle n_i \rangle_{\mathrm{GS}},
\end{equation}
where the index ``$\mathrm{an}$'' means that the density is evaluated in the presence of anyons, and ``$\mathrm{GS}$'' means the density evaluated in the ground state (i.e.\ without anyons). Note the difference from the definition used in Sec. \ref{ssec:GSParticleDensity} - now we do not care about the background charge, but only about the difference of the particle density distributions with and without anyons (otherwise we would always obtain some excess charge near the interface). The charge of the anyon is studied by investigating the charge accumulated within some radius around the anyon position $w_k$,
\begin{equation}
\tilde{Q}_k(r)= \sum_{i}\theta(r-|z_i-w_k|)\tilde{Q}_i.
\end{equation}
If the anyons are screened, $\tilde{Q}_k(r)$ should converge to a fixed value quickly as we increase $r$.

Fig.\ \ref{fig:AnyonDensityFigure} (a) shows the distribution of charge $\tilde{Q}_i$ in the case of two Abelian and two non-Abelian anyons on a cylinder.  Each of the anyons is located in a part where it is a valid topological excitation. It can be seen that they are indeed well screened, with most of the charge concentrated near their positions. The calculation of $\tilde{Q}_{k}(r)$, displayed in Fig.\ \ref{fig:AnyonDensityFigure} (b)-(e), shows that they indeed seem to converge to a value close to $-p_k/q$ as $r$ increases.

As noted in Sec. \ref{ssec:AnyonWfn}, the definition of our wavefunction does not forbid us to put the non-Abelian anyons within the $R$ part. The result of exchanging one Abelian and one non-Abelian anyon from Fig.\ \ref{fig:AnyonDensityFigure} (a) is shown in Fig.\ \ref{fig:AnyonDensityFigure} (f). It can be seen that still the charge is concentrated mostly in their vicinity, and approximately has the expected value $-p_k/q$. 

We note that in some cases, even with Abelian anyons only, there is some additional charge modulation at the interface. This is a finite-size effect, whose strength decreases with $N_y$. The detailed analysis of this effect can be found in Appendix \ref{app:anyons}. We note that a similar phenomenon was encountered for Laughlin/Laughlin interfaces \cite{jaworowski2020model}.

\begin{figure}
\includegraphics[width=0.5\textwidth]{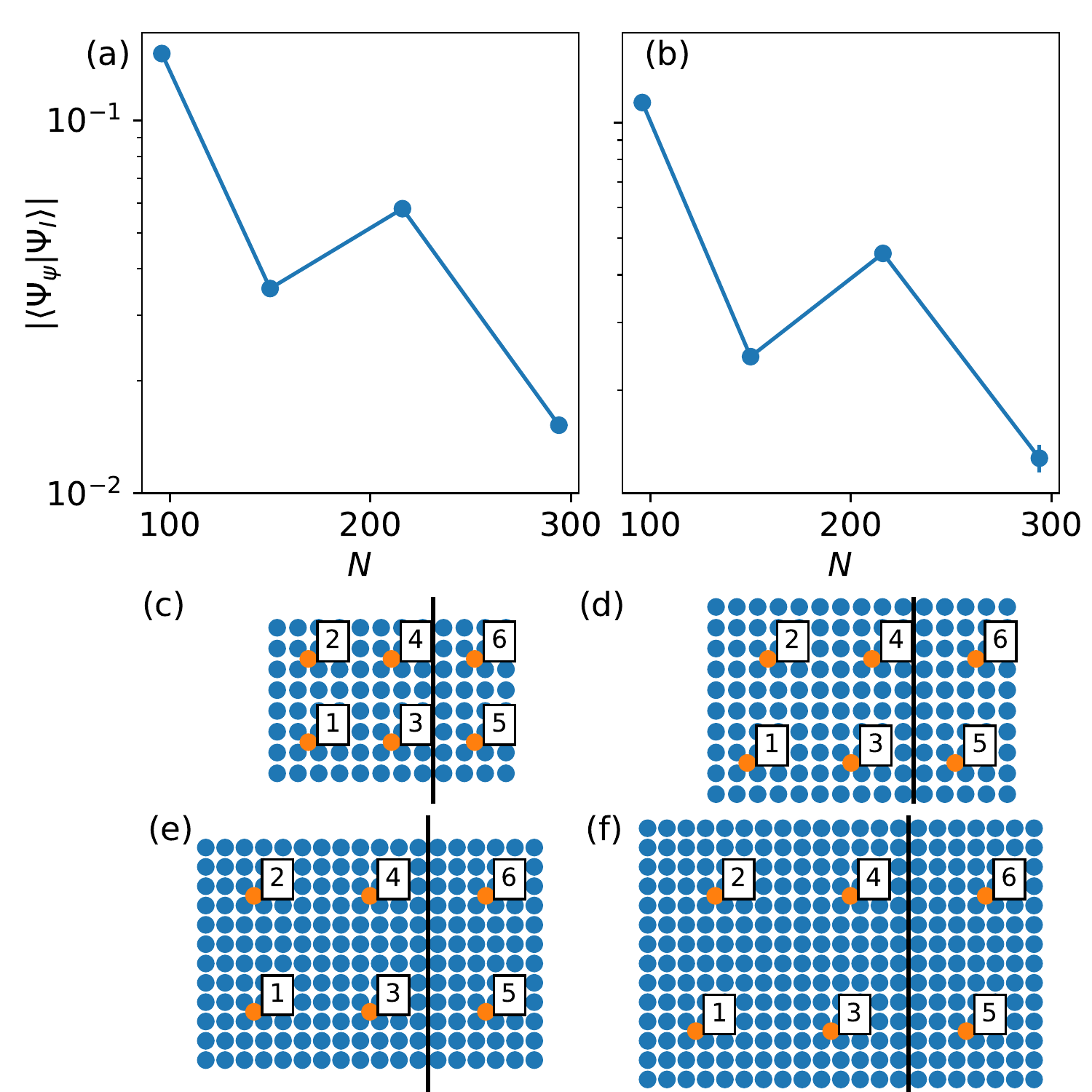}
\caption{
The overlap between conformal blocks as a function of the number of sites for the planar systems of size $(2k+k)\times 2k$, for $k=4,5,6,7$. Both subplots correspond to 4 non-Abelian anyons in the $L$ part and two Abelian ones in the $R$ part: (a) $p_1=p_2=-0.5$, $p_3=p_4=0.5$, $p_5=1=-p_6$, (b) $p_1=p_2=p_3=p_4=0.5$, $p_5=p_6=-1$. The rest of the plots, (c)-(f), show the systems taken into account in the calculation. The blue and orange points denote the sites and the anyons, respectively. The numbers denote the ordering of the anyons, which fixes the basis for the degenerate states via \eqref{eq:ising_4qh}.}
\label{fig:overlap_scaling}
\end{figure}

\subsection{Anyon statistics}
To check whether the ``anyons'' we investigate are true anyons, we have to evaluate their statistics. We will consider the processes in which a single mobile anyon $l$ encircles other, static anyons. The effect of anyon braiding is given by
\begin{equation}
\boldsymbol{\Psi}=\boldsymbol{\gamma}^{\mathrm{M}}\boldsymbol{\gamma}^{\mathrm{B}} \boldsymbol{\Psi}
\end{equation}
where $\boldsymbol{\Psi}$ is a vector of degenerate wavefunctions $\ket{\Psi_\alpha}$ for all possible values of $\alpha$, while $\boldsymbol{\gamma}^{\mathrm{M}}$ and $\boldsymbol{\gamma}^{\mathrm{B}}$ are the monodromy and Berry matrices. The monodromy matrix can be evaluated from the analytical continuation of the wavefunctions, while the Berry matrix can be written as $\boldsymbol{\gamma}^{\mathrm{B}}=\exp\left(i\boldsymbol{\theta}^{\mathrm{B}} \right)$, where the elements of $\boldsymbol{\theta}^{\mathrm{B}}$ are given by
\begin{equation}
\theta^{B}_{\alpha\beta}=i \oint_{P}\braket{\psi_\alpha|\frac{\partial}{\partial w_l}\psi_\beta}\mathrm{d}w_l +\mathrm{c.c.},
\label{eq:Berry1}
\end{equation}
where $P$ is the path of the $l$th anyon.

To proceed further, we need to show that the conformal blocks are orthogonal if there is more than one. The overlaps can be computed using Monte Carlo, as explained e.g. in Ref.\ \cite{manna2018nonabelian}. In Fig.\ \ref{fig:overlap_scaling} (a) and (b) we plot the overlap between the two conformal blocks for two cases of four non-Abelian anyons and two Abelian ones in four systems, depicted in Fig.\ \ref{fig:overlap_scaling} (c)-(f). A general trend of overlap decreasing with $N$ is seen, with $|\braket{\Psi_{\psi}|\Psi_I}|$ of the order $10^{-2}$ for the largest system. This shows that in large systems that can be studied using Monte Carlo the conformal blocks are already close to orthogonality, and we can expect that in the thermodynamic limit the orthogonality will be achieved. %In Fig.\ \ref{fig:overlap_scaling} (a), (b), we also show the the ratio of squared normalization constants in the two conformal blocks, also computed as explained in Ref.\ \cite{manna2018nonabelian}, which will be useful later.

It can be shown \cite{manna2018nonabelian} that, assuming the conformal blocks are orthogonal or there is just one, we can express the Berry phase \eqref{eq:Berry1} solely using the normalization constant
\begin{equation}
\theta^{B}_{\alpha\beta}=i \delta_{\alpha\beta} \oint_{P} \frac{1}{2C_\alpha^2} \frac{\partial}{\partial w_l}C_\alpha^2\mathrm{d}w_l+\mathrm{c.c.}
\label{eq:Berry1_2}
\end{equation}
\subsubsection{Abelian anyons}\label{sssec:abelian}
In the case when the $l$th anyon is Abelian, the Berry phase can be computed analytically, under the assumption that the anyons are well-screened (which is supported by the numerical results from Sec. \ref{ssec:AnyonDensity}). The partial derivative $\frac{\partial}{\partial w_l}$ in such a case does not act on the Ising part of the wavefunction. Thus, we can easily generalize the reasoning from Refs.\ \cite{rodriguez2015continuum}. Knowing the wavefunction coefficients \eqref{eq:AnyonWfnCoeffs}, we can evaluate the derivative
\begin{multline}
\frac{\partial C_\alpha^2}{\partial w_l}=
C_\alpha^2 \sum_{k} \frac{p_l\langle n_k \rangle}{w_l-z_k}-
C_\alpha^2 \sum_{k} \frac{p_l\eta_k}{q(w_l-z_k)}+\\+
C_\alpha^2 \sum_{k(\neq l)} \frac{p_lp_k}{q(w_l-w_k)}.
\label{eq:dCsq_w}
\end{multline}
Thus, for an anticlockwise path winding at most once around each site and each anyon, the Berry phase is 
\begin{multline}
\theta^{B}_{\alpha\beta}=\delta_{\alpha\beta} \left[
\frac{i}{2}\oint_P\sum_{k} \frac{p_l\langle n_k \rangle}{w_l-z_k} \mathrm{d}w_l + \mathrm{c.c.}
\right]+ \\
+2\pi\delta_{\alpha\beta}\sum_{k:z_k\in S} \frac{p_i\eta_k}{q} -
2\pi\delta_{\alpha\beta}\sum_{\substack{k : w_k\in S, \\ k\neq i}} \frac{p_ip_k}{q} 
\label{eq:Berry2}
\end{multline}
where $S$ is the region of space encircled by the path $P$. 

To deal with the first term of Eq.\ \eqref{eq:Berry2}, we note that we are in fact not interested in the phase $\theta^{B}_{\alpha\beta}$ itself, because it contains both the statistical contribution and the Aharonov-Bohm contribution, arising from the encircled sites. For simplicity, let us assume that we compute the mutual statistics of anyons $l$ and $m$. To get rid of the Aharonov-Bohm phase, we compute the difference of Berry phases with and without encircling anyon $m$. That is, we consider two cases: $\theta^{\mathrm{B, in}}_{\alpha\beta}$, when anyon $m$ is inside $S$, and $\theta^{\mathrm{B, out}}_{\alpha\beta}$, when it is outside, while the positions of all the other anyons are the same in both cases. We have
\begin{multline}
\theta^{\mathrm{B, in}}_{\alpha\beta}-\theta^{\mathrm{B, out}}_{\alpha\beta}=\\=
\delta_{\alpha\beta}
\left[
\frac{i}{2} \oint_P\sum_{k} \frac{p_l(\langle n_k \rangle_{\mathrm{in}}- \langle n_k \rangle_{\mathrm{out}})}{w_l-z_k} \mathrm{d}w_l + \mathrm{c.c.} \right]+\\
-2\pi \frac{p_lp_m}{q} 
\end{multline}
where $\langle n_k \rangle_{\mathrm{in}}- \langle n_k \rangle_{\mathrm{out}}$ are the particle densities in the two cases. Now, the assumption of screened anyons comes into play. If the anyons are screened and far from each other, the density difference is nonzero only in the vicinity of the two locations of anyon $m$ and is $w_l$-independent. Thus, it can be taken out of the integral. Then, applying the residue theorem, we obtain
\begin{multline}
\theta^{\mathrm{B, in}}_{\alpha\beta}-\theta^{\mathrm{B, out}}_{\alpha\beta}=
-2\pi  p_l\sum_{k (z_k\in S)} (\langle n_k \rangle_{\mathrm{in}}- \langle n_k \rangle_{\mathrm{out}})
-2\pi \frac{p_lp_m}{q} 
\end{multline}
We note that the sum of density differences within region $S$ is equal to the charge of the anyon $m$, i.e.\ $-p_m/q$. And thus, the Berry phase vanishes.

Hence, the effect of the braiding is given by the monodromy matrix, which is equal to
\begin{equation}
\gamma^{M}_{\alpha\beta}=\delta_{\alpha\beta}\exp\left( 2\pi i p_lp_m/q \right)
\end{equation}
This recovers the Laughlin anyon statistics. The expression is valid in the entire system, i.e.\ in the parts $L$ and $R$ and for paths crossing the interface, which is consistent with the fact that the Abelian anyons are valid topological excitations of both parts. We also note that the above reasoning is valid even when $p_m$ is fractional, i.e.\ it yields also the mutual statistics of Abelian and non-Abelian anyons, but only if the latter is static. As far as this condition is fulfilled, there is no problem with putting a non-Abelian anyon in part $R$. The problems arise when it moves, as we will see in the next subsection.

\begin{figure}
\includegraphics[width=0.5\textwidth]{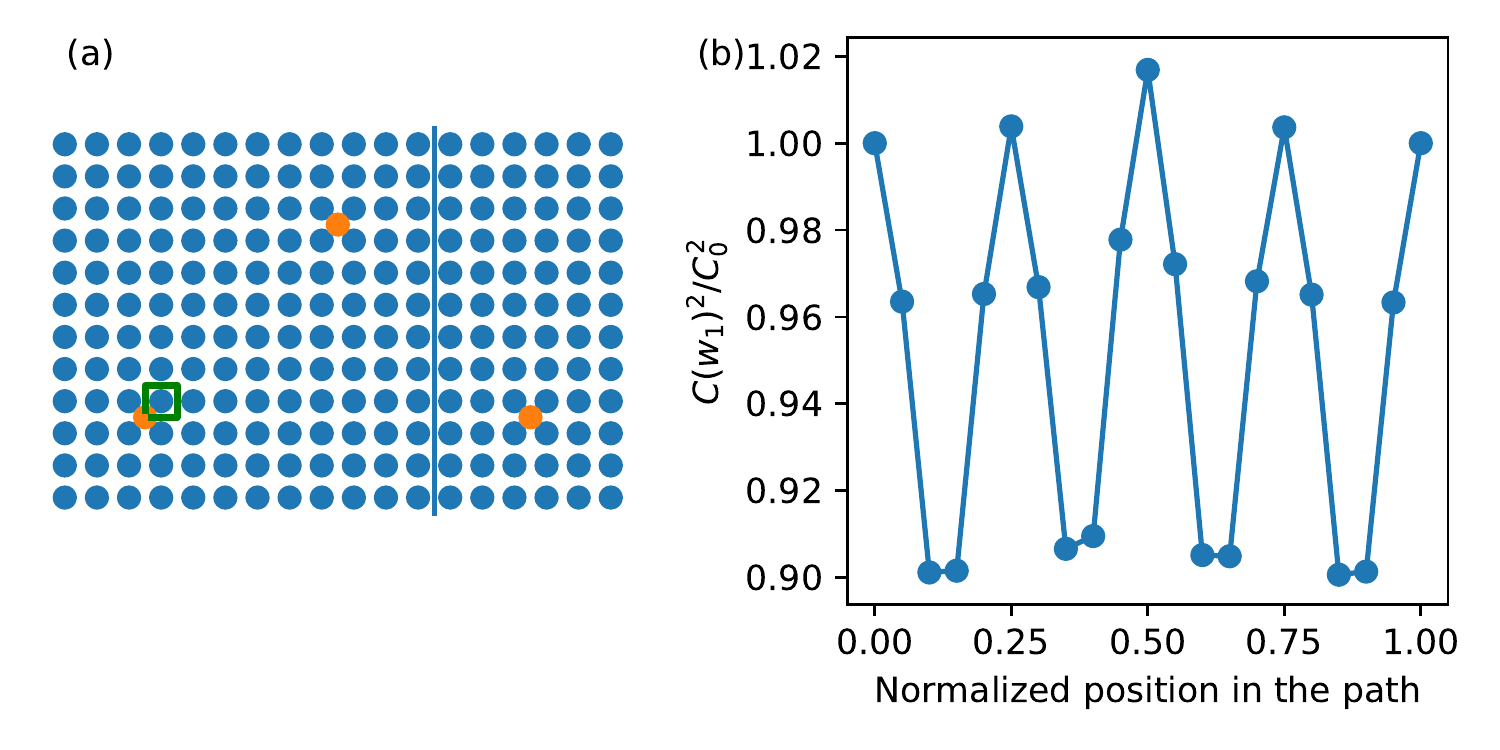}
\caption{(a), The path of the quasihole motion for a planar $(12+6) \times 12$ system with two non-Abelian quasiholes located in the $L$ part and one Abelian quasielectron located in the $R$ part. The orange dots mark the initial positions of the anyons, while the path is denoted by green lines. The quasihole moves anticlockwise along the path. (b) The corresponding ratio of squared normalization constants as a function of the quasihole position on the path.
}
\label{fig:OverlapFigure}
\end{figure}

\subsubsection{Non-Abelian anyons}\label{sssec:nonabelian}

In the case where a non-Abelian anyon is mobile, we verify the vanishing of the Berry phase numerically. Following Refs.\ \cite{manna2018nonabelian,manna2019quasielectrons}, we rely on the fact that the Berry phase vanishes if the normalization constant $C$ (and hence the integrand of \eqref{eq:Berry1}) is lattice-periodic in $w_l$ as long as anyon $l$ is far away from other anyons. To see this is the case, let us consider a planar system in which the anyon $l$ moves along a rectangular path consisting of four segments $P_1$, $P_2$, $P_3$, $P_4$. We consider $P_1$ and $P_3$ being parallel to the $x$ direction, with $x$ increasing in the former and decreasing in the latter, and located at $y=y_1$ and $y=y_2$. Similarly $P_2$ and $P_4$ are parallel to the $y$ direction, with $y$ increasing in the former and decreasing in the latter, and are located at $x=x_1$ and $x=x_2$. Moreover, we demand that the rectangle has integer dimension in the units of lattice constants, i.e.\ $x_2-x_1\in \mathbb{Z}$ and $y_2-y_1\in \mathbb{Z}$.  Then, we note that $\int_{P_1} f(w_l)\mathrm{d}w_l=\int_{x_1}^{x_2} f(x+iy_1)\mathrm{d}x$, and $\int_{P_3} f(w_l)\mathrm{d}w_l=\int_{x_2}^{x_1} f(x+iy_2)\mathrm{d}x$. If $f(w_l)$  is lattice-periodic, then $f(x+iy_1)=f(x+iy_2)$, and the contributions of $P_1$ and $P_3$ cancel each other. Similarly, one can show that the contribution of $P_4$ cancels the contribution of $P_2$. Thus, on this special path the statistics are determined by the monodromy. And, since the statistics are a topological property, we expect that they would not change if the path is deformed. The above reasoning can be regarded as a lattice generalization for the continuum argument that the Berry phase vanishes when $C$ is constant.

The lattice periodicity can be demonstrated by calculating the ratio of squared normalization constants in each point of some path, $C(w_l)^2/C_0^2$, where $C_0$ corresponds to the starting point of the path. The method of caluclating such ratios with Monte Carlo is described e.g. in \cite{manna2018nonabelian}. Because the system size required for the simulation of a complete braiding process is too large even for the Monte Carlo, we consider a square loop around a single lattice site (see Fig.\ \ref{fig:OverlapFigure} (a)), which will tell us how the normalization constant changes as we move an anyon by one lattice constant in the $x$ and $y$ directions (or both). 

We focus on the case of two non-Abelian anyons, for which there is only one conformal block.  We consider a $(12+6)\times 12$ planar system and arrange the anyons in the way shown in Fig.\ \ref{fig:OverlapFigure} (a). Fig.\ \ref{fig:OverlapFigure} (b) shows the resulting ratio of squared normalization constants while moving a quasihole around a small square of unit length. It can be seen that the dependence is nearly periodic, with ratio being close to 1 every time the anyon is at a corner of the square. The periodicity is not perfect -- there still are some discrepancies larger than the Monte Carlo error, which may be due to the insufficient separation of the mobile quasihole from other anyons or from the system edge.

Therefore, we expect that the statistical phase will be determined by the monodromy. We focus on the statistical contribution to the monodromy, i.e.\ the monodromy after removing the Aharonov-Bohm contribution  (which, in case of one anyon moving on a closed loop, can be computed by subtracting the phase with and without the second anyon within the path). This statistical contribution in the case of a single Moore-Read state is well-known \cite{bonderson2011plasma}. For a single anticlockwise exchange, it is equal to
\begin{equation}
\gamma^M=e^{i\pi (p_1p_2/q-1/8)}.
\label{eq:monodromy_2an}
\end{equation}
In the case of an interface, there are additional terms involving $R$ sites and non-Abelian anyons, but as long as the braiding path is located in the $L$ part, these terms do not contribute to the monodromy as no $R$ site is encircled by the anyon, thus the result of an exchange is still given by Eq.\ \eqref{eq:monodromy_2an}.

We can also ask what happens if we put the non-Abelian anyons in the $R$ part. In such a case, the monodromy indicates that the statistics become ill-defined. To see this, we note that now non-Abelian anyons encircle the $R$ sites. The factors $(w_l-z_i)^{p_ln_i}$ for $p_l=\pm 1/2$ introduce a nontrivial contribution to the monodromy every time $n_i\neq 0$, i.e.\ when a filled site is encircled. Thus, the monodromy depends on the path, i.e.\ it is not statistical. Moreover, since each configuration $\ket{\mathbf{n}}$ corresponds to different locations of the filled sites, each coefficient in \eqref{eq:AnyonWfn} transforms in a different way and thus the effect of a braiding is no longer a phase. Therefore, we conclude that it is not possible for non-Abelian anyons to cross the interface. This conclusion does not depend on the Berry phase (for some consideration regarding the Berry phase, see Appendix \ref{app:berry}).

%If we take four anyons, we have to fulfill three requirements: the orthogonality of the conformal blocks, the equality $C_\phi=C_I$ and the lattice periodicity of $C_\alpha$. Because the orthogonality is also important when calculating the statistics of Abelian anyons in presence of the non-Abelian ones, let us focus on it first. 
%
%????????????????????

We note that the factor $(w_l-z_i)^{p_ln_i}$ can generate a nonzero phase even if the non-Abelian anyon stays within the $L$ part, but moves around the cylinder, or encircles an $R$ island on the plane. However, in this case, the phase is well-defined as long as the parity of the number of encircled $R$ particles stays constant. For a single interface, this is guaranteed by the conservation of $L$ particle number parity -- which is another justification to our assumption of fermion parity conservation. The factor $(w_l-z_i)^{p_ln_i}$ acts also on $L$ particles going around the cylinder, but this does not generate inconsistency of the boundary conditions (see Appendix \ref{app:bc}). 

\begin{figure}
\includegraphics[width=0.5\textwidth]{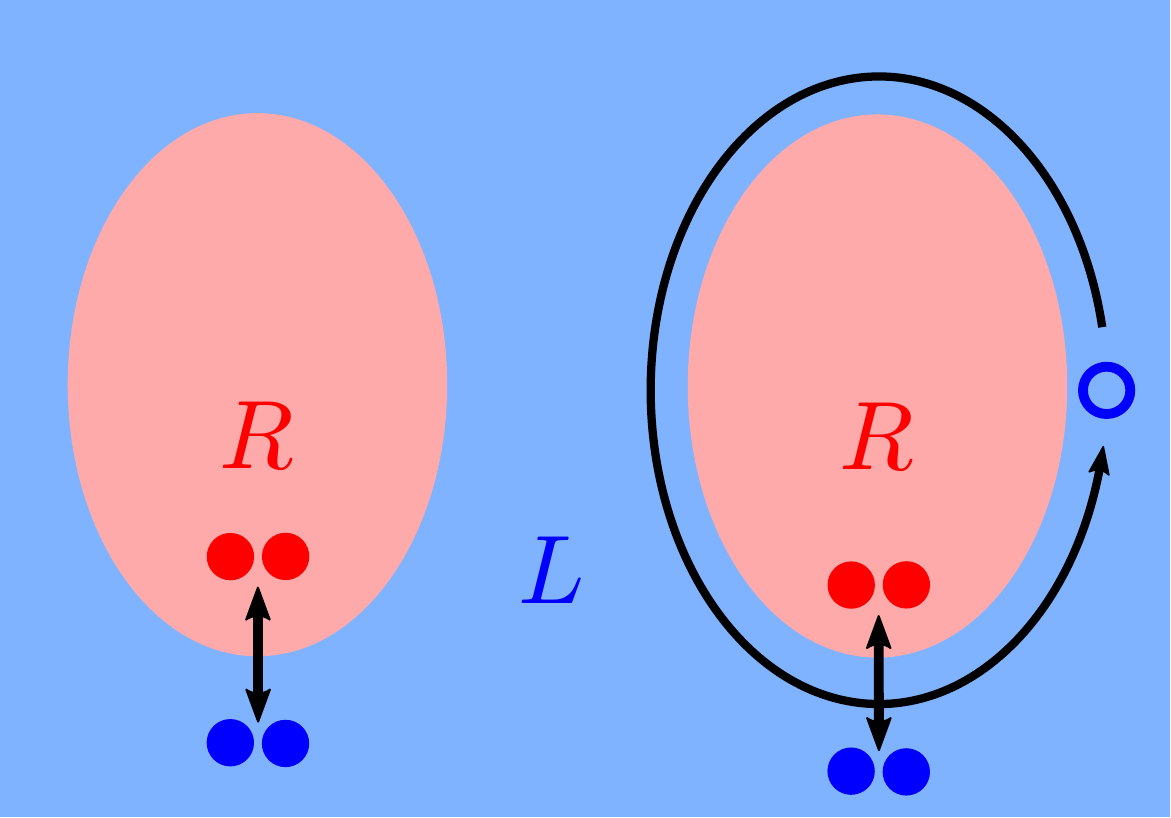}
\caption{A schematic depiction of a system with two $R$ islands on the $L$ plane. The filled and empty circles denote particles and anyons, respectively. The following processes are depicted here: the exchange of particles between $L$ and $R$ (the arrows at the bottom of each island) and measurement of the particle number parity (the arrow around the right island).}
\label{fig:degeneracy}
\end{figure}

\section{Multiple islands and topological degeneracy}\label{sec:islands}

So far, we discussed the properties of a single interface. Now, let us consider two disconnected islands of the $R$ type within an $L$ plane, as in Fig.\ \ref{fig:degeneracy}. Let us also assume that the processes of exchange of particles through the interfaces are local. That is, if the islands are sufficiently far apart from each other, a pair of particles annihilated from part $L$ should correspond to a creation of two particles in island 1 or two particles in island 2, but not one particle in each island, as shown in Fig.\ \ref{fig:degeneracy}. If we fix only the total number of particles $M$, the Hilbert space contains configurations where the numbers of particles in the first island $M_{R;1}$ is even and the ones where $M_{R;1}$ is odd. There is no local process connecting the two types of configurations. Therefore, the Hilbert space fragments into two disconnected subspaces. For each of them, we can define a model ground state wavefunction using Eq.\ \eqref{eq:pfafflin} or \eqref{eq:AnyonWfnCoeffs}, but reducing the basis only to the given subspace. If $k$ islands are introduced, then the Hilbert space fragments into $2^{k-1}$ subspaces, which is reminiscent of the degeneracy of the Majorana modes. The appearance of topological degeneracy in a similar setting of interfaces forming several disconnected islands was already discussed in Ref. \cite{bombin2011nested}. The $2^{k-1}$ topological degeneracy was also found for a Pfaffian/anti-Pfaffian interface \cite{lian2018theory}, and it is possible that the same mechanism can explain the topological degeneracy in our case.

The parity of $M_{R;1}$ can be measured by encircling a non-Abelian quasihole around it (see the arrow around the right island in Fig.\ \ref{fig:degeneracy}). Then, the factor $(w_l-z_i)^{p_ln_i}$ gives rise to a monodromy phase 0 if $M_{R;1}$ is even and $\pi$ if it is odd. 

In such a way, the $R$ islands can store quantum information even though the interfaces are gapless. We note that essentially the same mechanism of creating topological degeneracy can be applied to the Laughlin-Laughlin interfaces from Ref.\ \cite{jaworowski2020model}, thus connecting our model wavefunctions to earlier results, predicting the appearance of parafermion zero modes at some Laughlin-Laughlin interfaces \cite{santos2017parafermionic}.

\section{Conclusions}\label{sec:conclusions}
In this work, we have constructed model wavefunctions for lattice systems at filling $\nu=1/2$, in which part of the system is in the fermionic Moore-Read state, and the rest is in a bosonic Laughlin state. We considered the cases in the absence and presence of screened anyonic excitations.

We have seen that the conditions of reflection and scaling invariance lead to different lattice filling factors $\nu_{\mathrm{lat}}$, i.e.\ different particle densities on the two sides of the interface. %The accumulated charge grows with the total number of sites $N$, which raises doubts on the existence of the thermodynamic limit. Nevertheless, we can still consider finite-size systems (which may be more suited for optical lattice experiments anyway), 
For wide enough systems, these densities are nearly constant in the bulks of the two parts of the system, and their values are close to what is expected for the respective single quantum Hall states. Also, the constant term $\gamma$ of the entanglement entropy scaling in the bulks is consistent with the values characterizing the topological order of the respective quantum Hall states.

As for the interface itself, we have found that some charge accumulates in its vicinity, due to the fact that the particle density varies more smoothly than the background charge. We observed a lack of exponential decay of the correlation function in its vicinity, consistent with the prediction that the interface is gapless. We have also shown that for the investigated system sizes the scaling of the entanglement entropy at the interface is approximately linear, although the data are too noisy to determine the coefficient exactly.

We have studied the properties of the Laughlin anyons (which are valid topological excitations of the entire system) and the basic MR non-Abelian anyons. We have found that the quasiparticles of both types are well-screened and have the expected charge irrespective of their location. However, the statistics become ill-defined if the path of a non-Abelian anyon passes through part $R$.

Moreover, we argued that for multiple, disconnected islands of the $R$ part within an $L$ system, the particle number parity at each island cannot be changed locally, i.e.\ it is topologically protected. It can be measured by braiding a non-Abelian anyon around the island.

The presented construction can be modified and extended in several ways. First, after allowing double occupancy, one can consider an interface between a bosonic MR state and a fermionic integer quantum Hall state. Secondly, one can also consider different fillings $\nu$ on both sides, which would allow for all-bosonic or all-fermionic systems, at the price of enforcing different charges of the particles on the two sides. Finally, one can also use other quantum Hall states -- e.g. by forming a MR/Halperin interface, studied in \cite{yang2017interface,crepel2019variational} for the continuous case.

%Let us finally comment on the connection between our construction and experiment. The discrete nature of our system suggests that we should look for potential experimental realization in lattice quantum simulators, such as optical lattices. However, there are several major obstacles in this way: (i) we did not obtain a parent Hamiltonian, (ii) the tunneling between the $L$ and $R$ parts means that pairs of bosons are transmuted into pairs of fermions and vice versa, (iii) only pairs of particles can tunnel through the interface and (iv) even a single lattice quantum Hall state was not achieved in quantum simulators. It might be possible to solve problem (i) by performing exact diagonalization of small systems with a short-range Hamiltonian and optimizing its parameters to maximize overlap with our wavefunction.  As for (ii), while in the context of optical lattices the statistical transmutation was considered only in 1D \cite{keilmann2011statistically,greschner2015anyon}, it was proposed for 2D arrays of optical cavities \cite{zhu2018hadware} as a nonlocal transformation. Ad (iii): There is a lot of effort directed towards performing such experiments, so we hope that few-particle quantum Hall states will be achievable in near future.

\acknowledgements{This work was in part supported by the Independent Research Fund Denmark under grant number 8049-00074B. BJ was supported by START Fellowship by Foundation for Polish Science (FNP), no. 32.2019.}

\appendix

\begin{figure}
\includegraphics[width=0.5\textwidth]{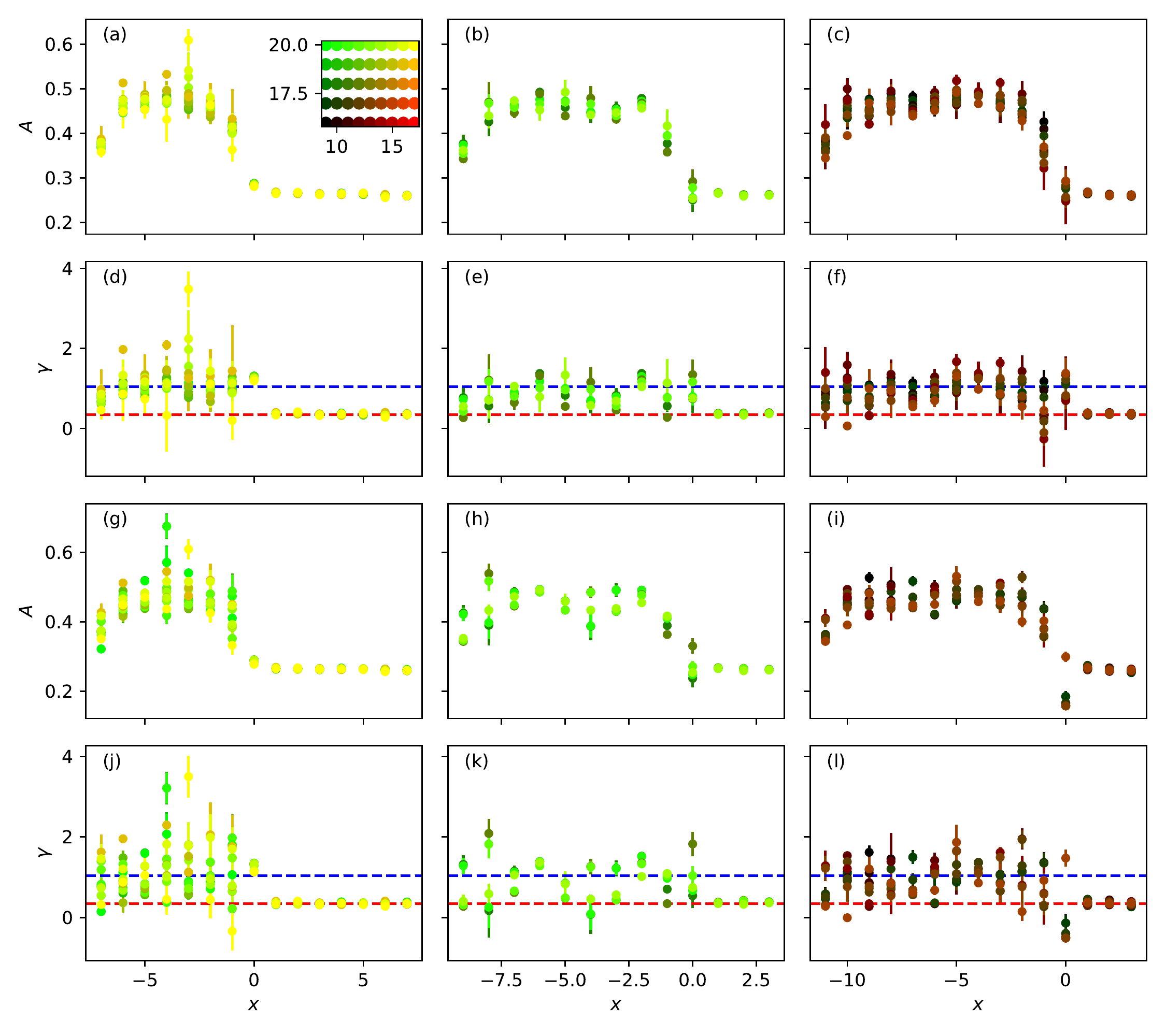}
\caption{The parameters of the linear fit of the entanglement entropy scaling as a function of $x$ position of the cut, for different sets of data points included. Columns 1,2,3 correspond to systems of size $(8+8)\times N_y$,  $(10+4)\times N_y$, $(12+4)\times N_y$, respectively. Rows 1 and 3 show the gradient $A$ of the fit, while rows 2 and 4 show $\gamma$, i.e.\ minus the intercept.  The results in the first (last) two rows were obtained without (with) the inclusion of error bars in the weights for the fit. The colors denote the different sets of data points included, according to the colormap in the inset of (a) (the horizontal and vertical axes correspond to $N_{y, \mathrm{min}}, N_{y, \mathrm{max}}$, respectively).
}
\label{fig:entropy_appendix}
\end{figure}

\section{Controlling the coupling across the interface}\label{app:copuling}
It is natural to expect that if the $L$ and $R$ parts are sufficiently separated from each other, they act as independent Moore-Read and Laughlin systems. This is indeed the case. Let us consider the planar system as in Fig.\ \ref{fig:systemview} (a), except that the coordinates of the $L$ and $R$ sites are $\tilde{z}_i=z_i$ and $\tilde{z}_i=z_i+\Delta$, respectively, where $\Delta$ is a real number, and $z_i$ are the original site positions on the perfect square lattice as in Fig.\ \ref{fig:systemview}(a). In other words, $\Delta$ controls the distance between the $L$ and $R$ parts, while the relative positions within each part are unchanged. We factor each wavefunction coefficient \eqref{eq:pfafflin} into four factors $\Psi(\mathbf{n})=\delta_{\mathbf{n}}\Psi_L(\mathbf{n}_L)\Psi_R(\mathbf{n}_R)\Psi_{LR}(\mathbf{n})$, with $\delta_{\mathbf{n}}$ defined by \eqref{eq:cn}, and
\begin{equation}
\Psi_L(\mathbf{n}_L)=\mathrm{Pf}\left(\frac{1}{z_i'-z_j'} \right)\prod_{j=1}^{N_L}\prod_{i=1}^{j-1}(z_i-z_j)^{qn_in_j-\eta_Ln_i-\eta_Ln_j},
\end{equation}
\begin{multline}
\Psi_R(\mathbf{n}_R)=\prod_{j=N_L+1}^{N}\prod_{i=N_L+1}^{j-1}(\tilde{z}_i-\tilde{z}_j)^{qn_in_j-\eta_Rn_i-\eta_Rn_j}=\\=
\prod_{j=N_L+1}^{N}\prod_{i=N_L+1}^{j-1}(z_i-z_j)^{qn_in_j-\eta_Rn_i-\eta_Rn_j},
\end{multline}
\begin{multline}
\Psi_{LR}(\mathbf{n})=\prod_{j=N_L+1}^{N}\prod_{i=1}^{N_L}(\tilde{z}_i-\tilde{z}_j)^{qn_in_j-\eta_Rn_i-\eta_Ln_j}=\\=
\prod_{j=N_L+1}^{N}\prod_{i=1}^{N_L}(z_i-z_j-\Delta)^{qn_in_j-\eta_Rn_i-\eta_Ln_j},
\end{multline}
where $\Psi_L(\mathbf{n}_L)$ and $\Psi_R(\mathbf{n}_R)$ are essentially the same as the wavefunction coefficients of single MR and Laughlin states, respectively, apart from the charge neutrality condition which allows the exchange of particles between the $L$ and $R$ parts.

When $\Delta$ is sufficiently large, it dominates over the $z_i$s, and we can write
\begin{multline}
\Psi_{LR}(\mathbf{n})\approx \prod_{j=N_L+1}^{N}\prod_{i=1}^{N_L}(-\Delta)^{qn_in_j-\eta_Ln_i-\eta_Ln_j}=\\=(-\Delta)^{qM_LM_R-N_L\eta_L M_R-N_R\eta_RM_L}=\\=
(-\Delta)^{(qM_L-N_L\eta_L)(q M_R-N_R\eta_R)/q -N_LN_R\eta_L\eta_R/q}.
\end{multline}
Due to \eqref{eq:cn} we have $qM_L+qM_R=N_L\eta_L+N_R\eta_R$ and thus the term $(qM_L-N_L\eta_L)(q M_R-N_R\eta_R)$ is either zero or negative. Therefore, for large $\Delta$ the highest-weight configurations are the ones in which it is zero (if there are any). These are the ones in which the charge neutrality conditions for single quantum Hall states, $q_IM_I=N_I\eta_I$, are satisfied. In the limit $\Delta\rightarrow\infty$ these are the only remaining configurations. As $\Psi_{LR}(\mathbf{n})=\mathrm{const}$ within these configurations in this limit, the $L$ and $R$ parts become independent from each other, and the wavefunction of the entire system can be understood as a tensor product of Laughlin and Moore-Read states.

If there are no states where $q_IM_I=N_I\eta_I$ (or this condition would force $M_L$ to be odd, which is not possible), then the highest-weight configurations have $M_L$ and $M_R$ such that they are closest to charge neutrality of respective quantum Hall states (i.e. maximize $(qM_L-N_L\eta_L)(q M_R-N_R\eta_R)$). If there is just one such choice of $M_L$, $M_R$, then at large $\Delta$ we a obtain a tensor product of MR and Laughlin wavefunctions without charge neutrality (i.e. with anyons not pinned to a certain point on the plane or with edge states \cite{glasser2016lattice}). It may, however, occur (as in the case $(N_L~\mathrm{mod}~8)=4$) that there are more such choices of $M_L$, $M_R$, and thus for $\Delta\rightarrow\infty$ the wavefunction is a superposition of such tensor products.

\section{More details on entanglement entropy results}\label{app:entropy}
Since the entanglement entropy scaling on the $L$ side and at the interface is noisy, here we provide additional results. In Fig.\ \ref{fig:entropy_appendix}, we provide the fit parameters $A,\gamma$ for all possible positions of the entanglement cut parallel to the interface. Moreover, we also study different sets of data points characterized by different bounds $N_{y, \mathrm{min}},N_{y, \mathrm{max}}$. Each color corresponds to a fit based on datapoints $N_{y, \mathrm{min}},N_{y, \mathrm{min}}+1,\dots, N_{y, \mathrm{max}}$. 

It can be seen that on the $L$ side and at the interface, the results display large fluctuations, sometimes larger than the result itself. Nevertheless, the obtained values seem to be consistent with the $\gamma_L=\ln(8)/2$ prediction. There is also no reason to suggest that at the interface or at its vicinity $\gamma$ has a different value than on the left.

On the other hand, in the $R$ part, the results are close to $\ln(2)/2$ for every choice of data points.

\begin{figure}
\includegraphics[width=0.5\textwidth]{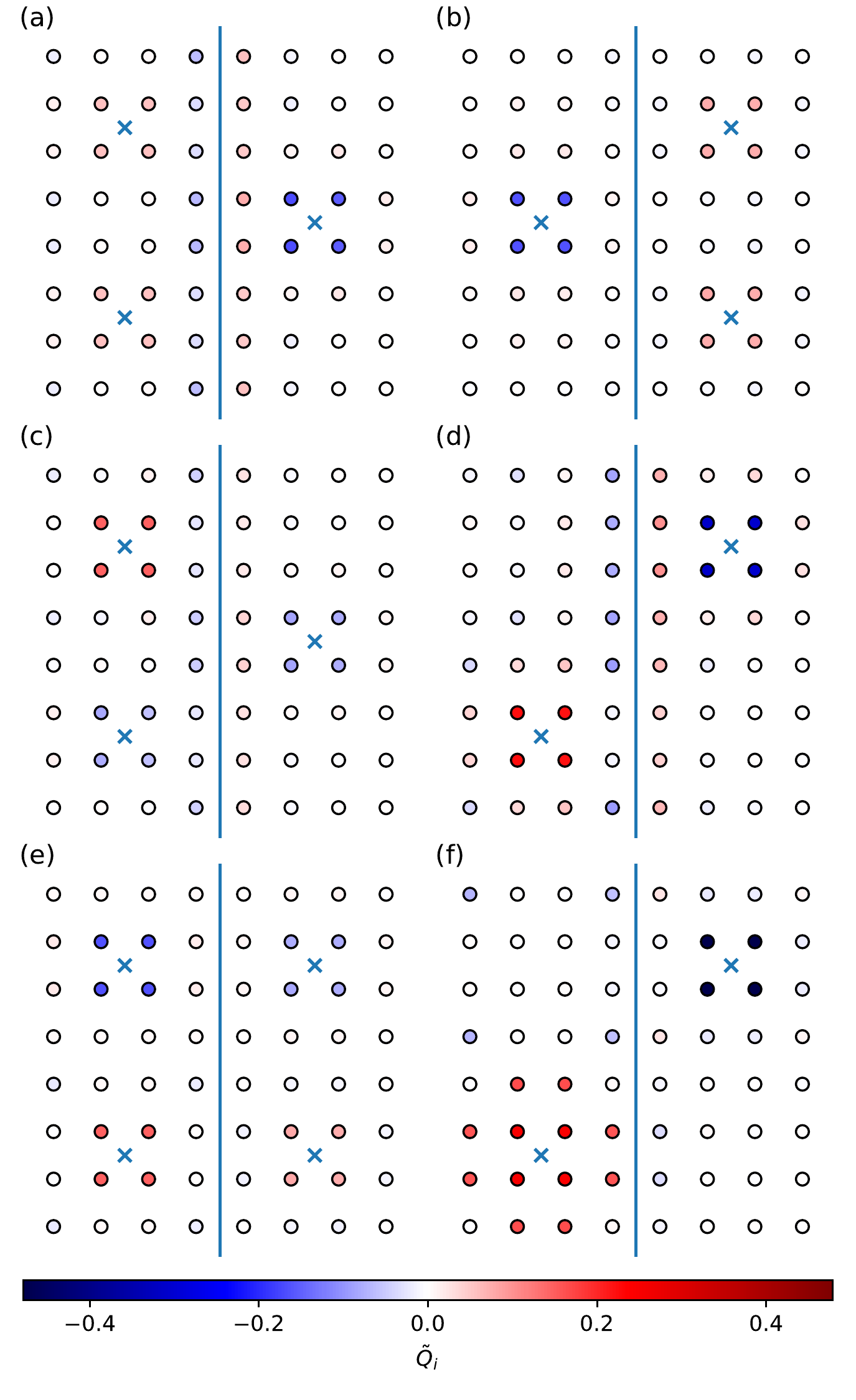}
\caption{The $\tilde{Q}_i$ charge density profiles in the presence of various anyon configurations for a $(4+4)\times 8$ cylinder: (a), (b) two non-Abelian quasielectrons and one Abelian quasihole with $p_i=1$, (c) two non-Abelian quasiholes and one Abelian quasielectron with $p_i=-1$, (d) an Abelian quasihole-quasielectron pair with $p_i=\pm 2$ (equivalent to an $R$ particle-hole pair), (e) an Abelian and a non-Abelian quasihole-quasielectron pair ($p_i=\pm 0.5$, $p_i=\pm 1$) (f) an Abelian quasihole-quasielectron pair with $p_i=\pm 4$ (equivalent to two holes and two particles). The anyon positions are marked with a ``$\times$'' symbol.
}
\label{fig:anyon_appendix1}
\end{figure}

\begin{figure}
\includegraphics[width=0.5\textwidth]{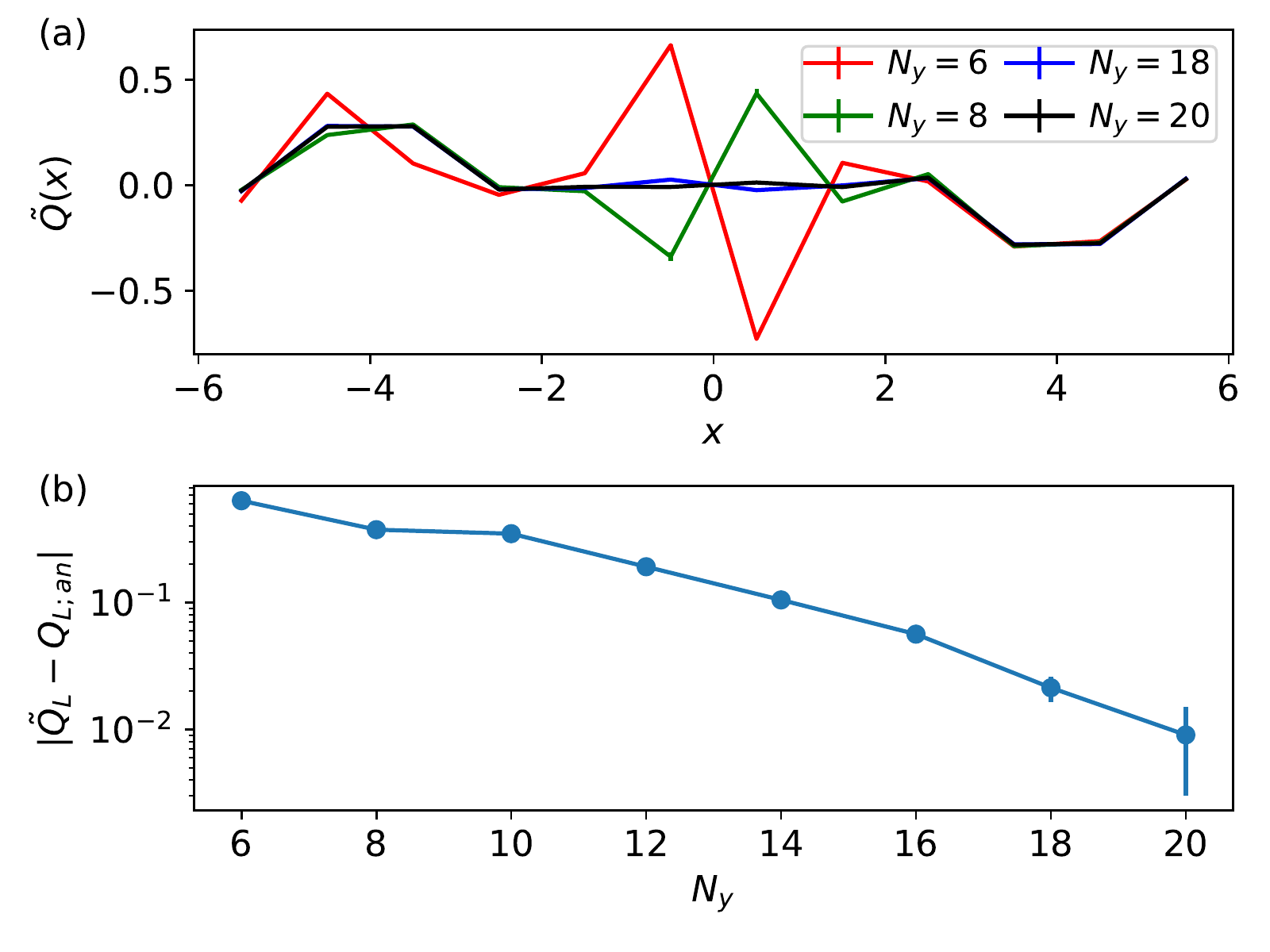}
\caption{The scaling of the density modulation at the interface for $(6+6)\times N_y$ cylindrical systems. (a) The density profile $\tilde{Q}(x)$ for relatively low and high $N_y$, even and odd. (b) The absolute value of the total charge in the $L$ part minus the anyon charge in the $L$ part as a function of $N_y$.
}
\label{fig:anyon_appendix2}
\end{figure}

\section{Density modulation at the interface in the presence of anyons}\label{app:anyons}
For some systems and some anyon configurations, we observe that $\tilde{Q}_i$ is nonzero also far from the anyon positions. Typically, the deviation is strongest near the interface, similarly to the Laughlin/Laughlin case \cite{jaworowski2020model}. In Fig.\ \ref{fig:anyon_appendix1}, we show the charge density $\tilde{Q}_i$ for six examples of anyon configurations for a $(4+4)\times 8$ cylinder. And indeed, for four of them some charge accumulates near the interface on both sides with the charge on the left being approximately equal to the charge on the right.

The presence and strength of these charge modulations depend on the anyon configuration, as well as the size of the system (especially $N_y$ and $(N_L~\mathrm{mod}~8)$). In Fig.\ \ref{fig:anyon_appendix2} (a) we plot the excess charge as a function of $x$,
\begin{equation}
\tilde{Q}(x)=\frac{\sum_{i}\tilde{Q}_i\delta(x_i-x)}{N_y}.
\end{equation}
for some $(6+6)\times N_y$ systems, with two non-Abelian quasielectrons placed at $x=-4$ and two non-Abelian quasiholes placed at $x=4$. The sign of the density modulation depends on the parity of $N_y$, which corresponds to the two possibilities $(N_L~\mathrm{mod}~8)=0,4$ (see the red and green curves in Fig.\ \ref{fig:anyon_appendix2} (a), denoting $N_y=6$ and $N_y=8$, respectively). The magnitude of the density variation decreases with $N_y$, and at $N_y=19$ or $N_y=20$ (blue and black curves in Fig.\ \ref{fig:anyon_appendix2} (a), respectively) it appears to be almost gone. 

To quantify the accumulated charge, we define the total excess charge in part $L$,
\begin{equation}
\tilde{Q}_{L}=\sum_{i\leq N_L} \tilde{Q}_{i},
\end{equation}
and the anyon charge in part $L$,
\begin{equation}
Q_{L; \mathrm{an}}=-\sum_{i: w_i\in L} p_i/q.
\end{equation}
In Fig.\ \ref{fig:anyon_appendix2} (b) we plot the $|\tilde{Q}_{L}-Q_{L; \mathrm{an}}|$, i.e.\ the magnitude of the excess charge in part $L$ not associated with anyons. It decreases exponentially with $N_y$, suggesting that for infinitely wide cylinders the only excess charge is concentrated in the vicinity of the anyon positions. The excess charge near the interface, $\tilde{Q}(x=-0.5)$, behaves very similarly to the plot in Fig.\ \ref{fig:anyon_appendix2} (b).

\begin{figure}
\includegraphics[width=0.5\textwidth]{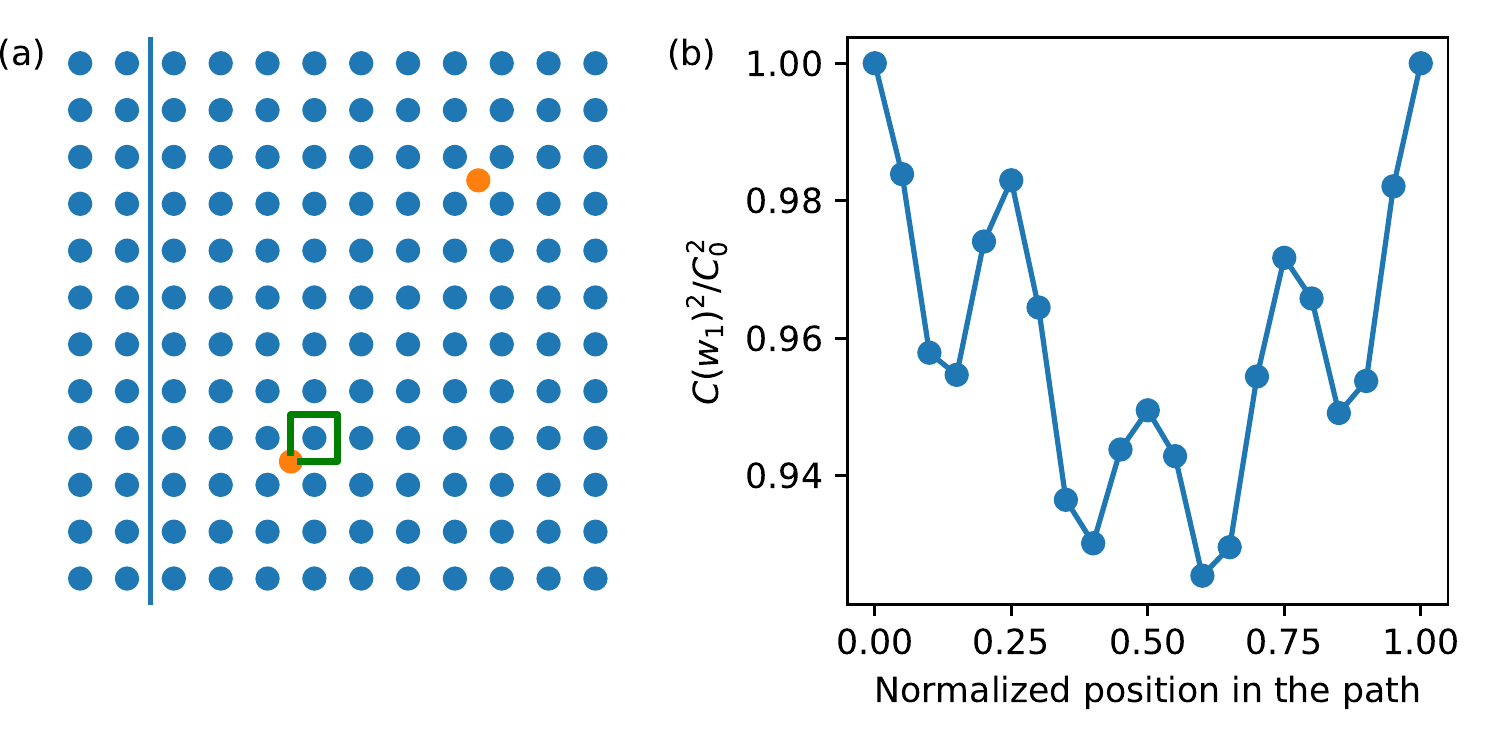}
\caption{(a) A path of the quasi-electron motion in a planar $(2+10)\times 12$ system. The orange points are the initial position of the anyons, while the green lines denote the anticlockwise path of the non-Abelian quasielectron motion. The second anyon is a non-Abelian quasihole. (b) The ratio of squared normalization constants in the given point on the path and in the initial point.}
\label{fig:app_berry}
\end{figure}

\section{Berry phase for non-Abelian anyons in the $R$ part}\label{app:berry}
We have shown that if we put the non-Abelian anyons in the $R$ part, the monodromy in the braiding process becomes ill-defined, and thus they lose their anyonic behaviour. For completeness, here we consider the Berry phase contribution.

Let us first consider a trivial case of $N_L=0$, with two non-Abelian quasiholes ($p_1=p_2=1/2$) and one Abelian quasielectron ($p_3=-1$).  In this case, the Pfaffian equals 1, and the only non-constant term in the Ising correlator, $(w_1-w_2)^{-1/8}$, is canceled by the $(w_1-w_2)^{p_1p_2/q}$ term from the Jastrow part. We consider moving quasihole 1 on a closed loop, and compare the cases where the quasihole 2 is inside and outside the path. Following the same approach as in Sec. \ref{sssec:abelian}, we arrive at $\theta^{\mathrm{B,in}}-\theta^{\mathrm{B,out}}=2\pi \frac{p_1p_2}{q}=\pi/4$.

The case of $N_L=2$ is a little more complicated, but still tractable analytically under the assumption of  screened anyons. Let us again focus on the case of two non-Abelian quasiholes ($p_1=p_2=1/2$) and one Abelian quasielectron ($p_3=-1$). The wavefunction is now given by
\begin{multline}
\Psi(\mathbf{w}, \mathbf{n})=2^{-\frac{n_1+n_2}{2}}A^{\frac{n_1+n_2}{2}} (w_1-w_3)^{p_1p_3/q}(w_2-w_3)^{p_2p_3/q}\times \\ \times
\prod_{i=1}^2 \prod_{j=3}^N (w_i-z_j)^{p_in_j}
\prod_{i=1}^{N}(w_3-z_i)^{p_3n_i} 
\prod_{i=1}^3\prod_{j=1}^{N}(w_i-z_j)^{-p_i\eta_j/q} \times \\ \times
\prod_{i<j}(z_i-z_j)^{qn_in_j}\prod_{i\neq j}(z_i-z_j)^{-n_i\eta_j}
\label{eq:wfnNL2}
\end{multline}
where 
\begin{equation}
A=\frac{(w_1-z_1)(w_2-z_2)+(w_1-z_2)(w_2-z_1)}{z_1-z_2}.
\end{equation}
In Eq.\ \eqref{eq:wfnNL2}, $A$ is raised to the power $\frac{n_1+n_2}{2}$, because there are only two options: either two $L$ sites are empty, and Pfaffian equals 1 ($\frac{n_1+n_2}{2}=0$), or they are both filled and the Pfaffian is $A$ ($\frac{n_1+n_2}{2}=1$).

The derivative of $\Psi$ is given by
\begin{multline}
\frac{\partial \Psi}{\partial w_1}=\frac{n_1+n_2}{2} \frac{\partial A}{\partial w_1}\frac{\Psi}{A}+
\sum_{i>2}\frac{ p_1 n_i}{w_1-z_i}\Psi+\\
-\sum_{i}\frac{ p_1 \eta_i}{q(w_1-z_i)}\Psi
+\frac{ p_1 p_3}{q(w_1-w_3)}\Psi,
\end{multline}
and thus, using Eq.\ \eqref{eq:Berry1_2}, we can write the Berry phase in a process where the first quasihole encircles the second one as
\begin{multline}
\theta^{\mathrm{B}}=
\left[\frac{i}{4}\oint_P\frac{\langle n_1 \rangle +\langle n_2 \rangle}{A} \frac{\partial A}{\partial w_1}\mathrm{d}w_1\right.+\\+
\left. \frac{i}{2}\oint_P\sum_{i>2}\frac{ p_1 \langle n_i \rangle}{w_1-z_i}\mathrm{d}w_1+\mathrm{c.c.}
\right]
+\sum_{i\in S}2\pi \frac{p_1 \eta_i}{q}-\\-
\delta_{w_3}2\pi \frac{p_1 p_3}{q},
\end{multline}
where $S$ is the region enclosed by the path $P$, and $\delta_{w_3}=0,1$ for $w_3$ inside or outside $S$, respectively.

Now, we need to subtract the Berry phase for the second quasihole inside and outside $S$. We assume that the quasielectron is outside $S$ for both cases. The result is 

\begin{multline}
\theta^{\mathrm{B,in}}-\theta^{\mathrm{B,out}}=\\=
\left[\frac{i}{4}\oint_P\frac{1}{A}\left(\langle n_1 \rangle_{\mathrm{in}} +\langle n_2 \rangle_{\mathrm{in}}-\langle n_1 \rangle_{\mathrm{out}} -\langle n_2 \rangle_{\mathrm{out}} \right) \frac{\partial A}{\partial w_1}\mathrm{d}w_1\right.+\\+
\left. \frac{i}{2}\oint_P\sum_{i>2}\frac{ p_1 (\langle n_i \rangle_{\mathrm{in}}-\langle n_i \rangle_{\mathrm{out}})}{w_1-z_i}\mathrm{d}w_1+\mathrm{c.c.}
\right]
\label{eq:berry_NL2}
\end{multline}
If the quasiholes are well-screened and far within the $L$ part, the particle density on the two $L$ sites is the same in the two cases. Therefore, the first term of Eq.\ \ref{eq:berry_NL2} vanishes. The second term can be dealt with using the reasoning from Sec. \ref{sssec:abelian}, resulting in the phase $\theta^{\mathrm{B,in}}-\theta^{\mathrm{B,out}}=2\pi \frac{p_1p_2}{q}=\pi/4$.

In the case of $N_L>2$, we can attempt to prove the vanishing of the Berry phase numerically, as in Sec. \ref{sssec:nonabelian}. However, we observe a lack of periodicity, as seen in Fig.\ \ref{fig:app_berry}. Therefore, our current approach does not allow us to extract the value of the Berry phase. This does not rule out an appearance of periodicity for systems too large to be studied using our Monte Carlo software. 

\begin{figure}
\includegraphics[width=0.45\textwidth]{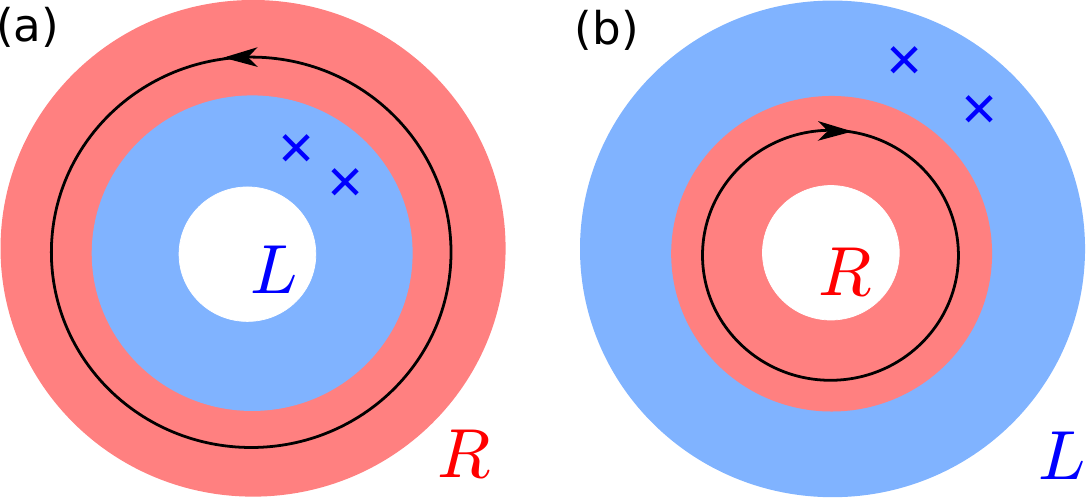}
\caption{Two ways of mapping the cylinder to the complex plane. The arrow denotes a path of $R$ particle, and the ``$\times$'' symbols denote the positions of the non-Abelian anyons.}
\label{fig:app_paths}
\end{figure}

\section{Boundary conditions for a cylinder with anyons}\label{app:bc}

The factor $(w_i-z_j)^{p_in_j}$, creating nontrivial monodromy when a non-Abelian anyon encircles a filled $R$ site, has a nontrivial effect also when an $R$ particle encircles a non-Abelian anyon. This can occur on a cylinder. The cylindrical system mapped to a complex plane looks as in Fig.\ \ref{fig:app_paths} (a). Thus, if an $R$ particle goes around a cylinder, it encircles the entire $L$ part, including the non-Abelian anyons located inside. But this can create a nonzero phase, which seems to lead to a conclusion that the boundary conditions in the $R$ part are determined by the number of anyons in the $L$ part. It would seem strange, as we can consider another, equivalent mapping of a cylinder to the complex plane, where the $R$ part is inside and no $L$ anyons are encircled, as in \ref{fig:app_paths} (b), and thus the boundary conditions for $R$ particles should be determined only by quantities related to the $R$ part.

Let us consider a path $K$ which encircles all $N_L$ $L$ sites, $k$ $R$ sites, as well as anyons of total charge $-P_{\mathrm{in}}/q$. The only factors which generate nonzero monodromy of an $L$ particle on path $K$ are $(z_i-z_j)^{-n_i\eta_j}$ and $(w_l-z_j)^{p_ln_j}$. They give rise to a phase

\begin{equation}
\phi=2\pi\left(P_{\mathrm{in}}-N_L\eta_L-k\eta_R\right).
\end{equation}
However, we can write it also as
\begin{multline}
\phi=2\pi\left(P-N_L\eta_L-N_R\eta_R\right)-\\
-2\pi\left(P_{\mathrm{out}}-(N_R-k)\eta_R\right),
\end{multline}
where $P_{\mathrm{out}}$ is minus the charge of all anyons outside the path in the units of $-1/q$ (i.e.\ their charge is $-P_{\mathrm{out}}/q$), and $P=P_{\mathrm{in}}+P_{\mathrm{out}}$. Due to the charge neutrality \eqref{eq:AnyonCN}, the first term vanishes. The second term depends only on the sites and anyons outside the path. Thus, it is possible to express the phase on path $K$ using only the quantities related to part $R$. In this way, one can see that the two mappings of the cylinder to the plane yield the same result. It is straightforward to generalize this reasoning to multiple interfaces.

%
%\section{The statistics of two non-Abelian anyons in the Abelian part in the case of $N_L=2$}\label{app:statistics}
%
%Here, we are going to show a simple example in which the the Berry phase of the non-Abelian anyons within the $R$ part is zero. 
%
%Let us consider a system with only two sites ($i=1,2$) in part $L$. For simplicity, we assume that the position of both are real and negative. To this system, we introduce two non-Abelian quasiholes and one non-Abelian quasielectron. We assume that the first quasihole is mobile, while the second one is located at $w_2=0$. The sites are arranged in such a way that the origin of the complex plane is located within part $L$. The quasielectron is located somewhere in part $L$. 
%
%The wavefunction in this case is given by
%\begin{multline}
%\psi(w, n)=A(w_1,w_2,z_1,z_2)(w_1-w_3)^{p_ip_3} \prod_{i>2}(w_1-z_i)^{p_1n_i}B(w_2,w_3, n_2,\dots,n_N)
%\end{multline}
%if $M_L=2$ and
%\begin{multline}
%\psi(w, n)=(w_1-w_3)^{p_ip_3} \prod_{i>2}(w_1-z_i)^{p_1n_i}B(w_2,w_3, n_2,\dots,n_N)
%\end{multline}
%if $M_L=0$. Here
%\begin{equation}
%A=\frac{(w_1-z_1)(w_2-z_2)+(w_1-z_2)(w_2-z_1)}{z_1-z_2}
%\end{equation}
%is the Pfaffian in the case of $M_A=2$ (i.e.\ the off-diagonal element of a $2\times 2$ matrix $\mathbf{A}$).

\bibliography{interfaces,quasiholes}
\end{document}